\documentclass[toc]{PoS}

\usepackage{amsmath} 
\usepackage[utf8]{inputenc} 
\usepackage{color,soul}
\setul{0.5ex}{0.3ex}
\setulcolor{blue}
\usepackage{slashed}
 \usepackage{bbm}
\usepackage{tikz}
\usetikzlibrary{matrix,arrows,positioning,shapes,snakes,fadings,decorations.pathmorphing,
decorations.pathreplacing,automata,shadows,decorations.markings,patterns}
\usepackage[thicklines]{cancel}
\usetikzlibrary{calc,shapes.callouts,shapes.arrows,cd}
 \usepackage[utf8]{inputenc}
 \usepackage[T1]{fontenc} 
 \usepackage{graphicx}
   \usepackage{subcaption}
\usepackage{changepage}
\newcommand{\Tr}{\operatorname{Tr}}
\newcommand{\be}{\begin{equation}}
\newcommand{\ee}{\end{equation}}

\title{An Introduction to Generalised Dualities and their Applications to Holography and Integrability}

\ShortTitle{Applications of Generalised Dualities}

\author{\speaker{Daniel C. Thompson} \\
    Department of Physics, Swansea University, Singleton Park, Swansea SA2 8PP, U.K.\\
    {\em and} Theoretische Natuurkunde, Vrije Universiteit Brussel   \& The International Solvay Institutes\\
Pleinlaan 2, B-1050 Brussels, Belgium \\  
        E-mail: \email{d.c.thompson@swansea.ac.uk}}

\abstract{These pedagogical lectures given at the  Corfu Summer Institute 2018 review two generalised notions of T-duality, non-Abelian T-duality and Poisson-Lie duality, and their applications. We explain how each of these has seen recent application in the context of holography. Non-Abelian T-duality has been used to construct new holographic dual geometries.  Poisson-Lie duality has been used to construct new integrable string sigma-models including the $\eta$- and $\lambda$-deformations of the $AdS_5\times S^5$ superstring  thought to encode quantum group deformations of holography. We also comment on the doubled worldsheet description that makes such dualities manifest.   }

\FullConference{Corfu Summer Institute 2018 "School and Workshops on Elementary Particle Physics and Gravity"\\
		(CORFU2018)\\
		31 August - 28 September, 2018\\
		Corfu, Greece}

\begin{document}
\section{Introduction}

Whilst the audience of this workshop consists primarily of string theorists, holographers and supergravity experts it is important to take a step back and appreciate the universal importance of duality.   Dualities have been a historic driver of theoretical progress with impact not only in string theory where they underpin  M-theory \cite{Hull:1994ys,Witten:1995ex}, but also in electromagnetism starting with the introduction of the monopole \cite{Dirac:1931kp} and then in gauge theory \cite{Montonen:1977sn,Osborn:1979tq,Seiberg:1994rs,Seiberg:1994pq},  statistical physics \cite{Kramers:1941kn,Savit:1979ny}  and condensed matter physics \cite{Luther:1975wr,Peskin:1977kp,Dasgupta:1981zz,Fradkin:1994tt} where they continue to provide stimulus \cite{Son:2015xqa,Karch:2016sxi,Seiberg:2016gmd}. 

An interesting  and particularly well understood example is found in two-dimensions.  Bosonisation \cite{Coleman:1974bu,Mandelstam:1975hb} gives the equivalence of a theory of bosons (e.g. the Sine-Gordon model) with that of a theory constituted by fermions (the Thirring model).  In these theories there is a $U(1)$ global symmetry and the duality transformation serves to interchange Bianchi identities   with conservation equations:  
\begin{equation}
 J_\mu \equiv \bar\psi \gamma_\mu \psi = \epsilon_{\mu \nu} \partial^\nu\phi \, . 
\end{equation}
Bosonisation  extends far beyond the simple example described above.  Witten \cite{Witten:1983ar} famously showed that a system of $N$ massless Majorana fermions $\psi^a$ with an $O(N)_L\times O(N)_R$ symmetry is dual (equivalent) to a bosonic theory whose variables are fields, $g$, taking values in a group manifold $G= SO(N)$.   The map can be made precise with the chirally conserved currents $J_\pm$ generating the $O(N)_L$ and $O(N)_R$ global symmetries obeying   duality rules 
  \begin{equation}
  J_+ =    g^{-1} \partial_+ g \ , \quad J_{-} =-   \partial_- g g^{-1} \ . 
\end{equation}
The chiral conservation in the dual bosonised picture corresponds exactly to the equations of motion of the Wess-Zumino-Witten (WZW) model.  Some years after bosonisation was established, a constructive path integral derivation was provided for both Abelian \cite{Burgess:1993np} and non-Abelian \cite{Burgess:1994np} cases.  This derivation works by gauging the global symmetry of the fermionic model and  introducing  Lagrange multiplier fields that ensure the corresponding gauge field strength is zero.  Performing the functional integral over the fermions and the gauge fields results in a bosonised theory whose degrees of freedom are, essentially, contained in the Lagrange multiplier fields.   

We now turn to string theory in which T-duality is the statement that, from the point of view of string theory, apparently different target spaces can describe the same underlying physics. In its simplest form the duality states an equivalence between the closed string theory whose target space has a compact $S_\theta^1$ of radius $R$ and that defined on a target space with a compact $S_{\tilde{\theta}}^1$ of radius $R^{-1}$.    
In many ways this is similar to the case of Abelian bosonisation described above.  First the duality serves to interchange a conservation law, that associated to the $U(1)$ global symmetry of the compact boson $\theta$ of radius R, with the Bianchi identity for the dual boson $\tilde{\theta}$: 
\begin{equation}
 J_\mu \equiv R \partial_\mu \theta =   \epsilon_{\mu \nu} \partial^\nu \tilde{\theta} \, . 
\end{equation} 
Just as with bosonisation the connection between the two variables involves derivatives, i.e. the map between variables is a non-local one.  Also akin to the Abelian bosonisation, the dual theory can be established through an analogous path integral manipulation\footnote{For narrative's sake here I am antichronological; the path integral derivation was of course established first for T-duality in the seminal work of Buscher then later exploited for Bosonisation and indeed other dualisations.}   known as the Buscher procedure \cite{Buscher:1987sk,Buscher:1987qj}.  

Given the close analogy between T-duality and Abelian bosonisation one might wonder if there is a T-duality equivalent to non-Abelian bosonisation.  More specifically, in string theory we are immediately prompted to ask if there is an extension of T-duality to the case of a non-Abelian set of isometries, or even if there are  still more exotic notions of T-duality?  A secondary question is then, if such dualities exist, what are they useful for?  The remainder of these lectures seek to outline some answers to these two questions.  

Our journey will involve exploring a hierarchy of possible dualities in the following string theory scenarios:
\begin{enumerate}
\item {\bf Abelian T-duality \cite{Kikkawa:1984cp,Sakai:1985cs,Buscher:1987sk,Buscher:1987qj}}.  Here there is a $U(1)^d$ action generated by Killing vectors, in adapted coordinates, $V_a = \partial_{\theta^a}$ which commute $  [L_{V_a} ,L_{V_b}]= 0$. That these correspond to isometries of the target space   implies the existence of a conserved Noether current $d\star J = 0$.    
\item {\bf Non-Abelian T-duality \cite{delaOssa:1992vci}}. Here there is a non-Abelian group $G$ action generated by Killing vectors  $V_a = v_a{}^i \partial_{x^i}$ which furnish the algebra $\frak{g}$ via  $[L_{V_a} ,L_{V_b}]= f_{ab}{}^cL_{V_b} $. That these correspond to isometries of the target space implies the existence of a conserved Noether current $d\star J_a = 0$.  
\item {\bf Poisson-Lie T-duality \cite{Klimcik:1995ux,Klimcik:1995dy}}. Here there is a non-Abelian group $G$ generated by  vectors  $V_a = v_a{}^i \partial_{x^i}$ which furnish the algebra $\frak{g}$ via  $[L_{V_a} ,L_{V_b}]= f_{ab}{}^{c}L_{V_c} $.  These are {\em not} required to be isometries of target space and so the conservation of the corresponding current is spoilt. However imposing a special structure on the target space allows the current to obey instead a modified conservation law   $d\star J_a =  \tilde{f}^{bc}{}_a\star J_b \wedge \star J_c  $ in which  $\tilde{f}^{bc}{}_a$ are structure constants of a dual algebra $\tilde{\frak{g}}$.  Remarkably, not only is it possible to find geometries for which this property holds, in this delicate situation one still has a notion of a T-dual sigma model, whose target space is $\tilde{G} =\exp \tilde{\frak{g}}  $.   
\end{enumerate}  

As is evident from the bibliography, these ideas are far from new and so one might ask why  they are relevant topics for this meeting; what has prompted a recent surge of activity in this area and indeed why had the topic remained dormant for many years in the first place?

  For a long time the non-Abelian and Poisson-Lie generalised notions of these ``dualities'' were viewed as exotica involving sigma-models with apparently rather intractable target space geometries.  This situation was exacerbated by their  unclear  status as fully fledged quantum dualities of string theory \cite{Giveon:1993ai}. Unlike Abelian T-dualities \cite{Rocek:1991ps} it, in general, can't be demonstrated that these classical canonical equivalences \cite{Curtright:1994be,Lozano:1995jx,Klimcik:1995dy,Sfetsos:1996xj,Sfetsos:1997pi} extend to the quantum theory in either of the expansion parameters of string theory.  Whilst robust to one-loop in worldsheet quantum corrections (i.e. $\alpha^\prime$ corrections), less is known definitively about higher orders.  It could be that the duality persists but the transformation of the background fields receives corrections order by order in   $\alpha^\prime$; though this remains unproven.  The second expansion, the string genus $g_s$ expansion, is afflicted with more challenges that we will comment on later.  It is fair to say the understanding of these issues still stands incomplete today.  

However with the advent of holography \cite{Maldacena:1997re} these topics have new life.  Indeed, in the best understood regime where the supergravity description is valid and at large $N$  one can, to a certain extent, set aside the important issues of $\alpha^\prime$ and $g_s$ corrections and ask if these generalised dualities can be employed usefully. 

 There are two key directions here. The first is that non-Abelian T-duality can be understood as a solution generating technique at the level of supergravity  \cite{Sfetsos:2010uq,Lozano:2011kb}.   This then has given a new tool in the arsenal of holographers to construct novel spacetimes that one hopes to be informative about (potentially non-Lagrangian) strongly coupled quantum field theories  starting with \cite{Lozano:2012au,Itsios:2012zv,Lozano:2013oma,Itsios:2013wd} (and see \cite{Lozano:2016kum} and references within for developments in this direction).  This direction is not without challenges, at present it seems we have only an incomplete knowledge of the global properties of the geometries produced in this way -- though recent work  \cite{Lozano:2016kum}  points to a way forward.  

The second direction concerns Poisson-Lie duality.  Though the geometries involved in Poisson-Lie (PL) duality  typically  have little or no apparent isometries (symmetries), this inelegance is illusory. As we will see later, when formulated in a way that makes the duality manifest the geometries are actually extremely simple and algebraically controllable.  Not only that, in a remarkable breakthrough,  Klim\v{c}\'{i}k  \cite{Klimcik:2002zj,Klimcik:2008eq} showed in certain circumstances PL $\sigma$-models can have infinite hidden symmetry and can be integrable.   In this lecture we shall call such integrable PL models {\em $\eta$-deformations} (they are also called Yang-Baxter $\sigma$-models) since they can be viewed as a deformation, labelled by a parameter $\eta \in \mathbb{R}$, of well understood integrable two-dimensional non-linear $\sigma$-models.   This idea gained significant traction following the seminal work of  Delduc, Magro and Vicedo \cite{Delduc:2013qra,Delduc:2014kha} with its extension to provide integrable deformations of the $AdS_5\times S^5$ superstring.  In this context it opens up the enticing possibility of a deformation of holography in which symmetry, and supersymmetry, are apparently reduced but the analytic tractability given by integrability is retained.  	
The $\lambda$-deformations introduced by Sfetsos \cite{Sfetsos:2013wia} provide another critical development that we will review in these lectures. These are integrable two-dimensional QFTs obtained as perturbations of Wess-Zumino-Witten (or gauged WZW) CFTs and at first sight seem rather disconnected from the aspects mentioned above.   However, as the eponymous parameter $\lambda$ that  defines these theories approaches unity, one recovers a non-Abelian T-dual geometry. Moreover, the $\eta$-deformations and the $\lambda$-deformations are related via a Poisson-Lie T-dualisation together with an analytic continuation \cite{Hoare:2015gda,Sfetsos:2015nya,Klimcik:2015gba}. The $\lambda$-deformation was extended to the $AdS_5\times S^5$ superstring by Hollowood, Miramontes and Schmidtt \cite{Hollowood:2014qma,Hollowood:2014rla} providing a Lagrangian description of a quantum group at root-of-unity deformation \cite{Hollowood:2015dpa}. 

Through these developments over the past few years we have now a wide story in which generalised notions of T-duality have been put to useful use. 
 \vskip 1cm 
{\em I have broadly reproduced the lectures given at the Corfu Summer Institute 2018 here, but for completeness I have chosen to supplement the notes with a few topics  and details that were not covered. This additional material is contained in sections indicated by an asterisk and can be skipped for a first orientation.    }

\section{Non-Abelian T-duality}

\subsection{The Principal Chiral Model and integrability} 
The main setting of these lectures will be on the string worldsheet.  We will be interested in strings propagating in a target space manifold ${\cal M}$, on which we introduce local coordinates $X^i$, $i = 1,\dots, d = \dim {\cal M}$,  equipped with a metric $G=G_{ij}dX^i \otimes dX^j$, a closed three-form $H= \frac{1}{6}H_{ijk} dX^i \wedge dX^j \wedge dX^k$ and a scalar field, the dilaton, $\Phi$.   Locally where we introduce a potential\footnote{Globally of course this may not be possible and indeed the WZW models discussed later will include this scenario.} such that $H=dB$, the dynamics is encoded by the non-linear $\sigma$-model action 
\be\label{eq:sigmamodel}
S= \frac{1}{\pi \alpha^\prime }\int_\Sigma  \mathrm{d}^2 \sigma  \, \partial_+ X^i \left( G_{ij}(X) + B_{ij}(X) \right) \partial_- X^j \ ,
\ee
in which $X^i$ are viewed as maps from the worldsheet, $\Sigma$, to the target space, $X:\Sigma \to {\cal M}$, and we adopt light-cone coordinates $\partial_\pm = \frac{1}{2} \left( \partial_\tau \pm \partial_\sigma\right)$ and use $d^2\sigma = d\sigma d\tau$. 

Our first goal is to understand T-duality in the case where the target space admits an action of a non-Abelian group $G$ (algebra $\frak{g}$) that generates isometries.  There are many ways to do this but the simplest example  is when the target space is the group manifold ${\cal M}=G$.  To formulate the theory in this case it is convenient to work with group valued maps $g: \Sigma \to {\cal M}$ rather than the local coordinates $X^i$, though of course we can always change back to such a description since  $g = g(X)$.  The simplest example of this type, in which we set the B-field to zero, is known for historical reasons as the {\em Principal Chiral Model} (PCM) and is described by the action 
\be\label{eq:PCM}
S_{PCM,\kappa}[g] = -\frac{ \kappa^2}{4\pi}\int_\Sigma  \mathrm{d}^2 \sigma  \, \Tr\left(g^{-1}\partial_+ g g^{-1} \partial_- g \right)    \ .
\ee
 This is a very special theory that enjoys a number of features:
 \begin{itemize}
 \item There is a global $G_L \times G_R$ symmetry that acts via $g \mapsto h_L\, g \, h_R^{-1}$ for $h_{L, R}\in G$. 
 \item The overall coupling $\kappa$ has a beta-function.  The theory is asymptotically free (in the UV the effective geometry flattens out) whilst in the IR  the coupling becomes strong.  At some scale the effective curvature becomes of string scale and the perturbative calculation ceases to be reliable; instead the theory dynamically generates a mass scale.   In this sense it is a $2d$ prototype to QCD.\footnote{One might object here, on the grounds that there are no instantons in the theory since $\pi_2(G)= 0$, however there are non-perturbative {\em uniton} configurations  \cite{Uhlenbeck} which have been conjectured to play a rather analogous role in the resurgent structure of the quantum theory \cite{Cherman:2013yfa}. For applications of the ideas of resurgence to the $\eta$-deformations, one of the key topics of these lectures, see \cite{Demulder:2016mja}.}   
 \item Whilst this is not a CFT, it can appear as a subsector of consistent string theory backgrounds.  For instance one can take $G=SU(2)$, for which ${\cal M} = S^3$, and find that the PCM  is part of the e.g. $AdS_3\times S^3 \times T^4$ type IIB superstring with the geometry curvature supported by RR flux.
 \item Most importantly for us is that the theory is {\em integrable} at the classical and indeed at the quantum level.   
 \end{itemize}    

Integrability will be a major theme in these lectures so let us review it.  For the most part I will use integrability in the classical sense (though see comments later on quantum integrability).  In the current context this means that the dynamics can be formulated in terms of a flat {\em Lax connection}.   Let us introduce currents 
\be
{\mathcal J}_\pm = g^{-1} \partial_\pm g \ , 
\ee
in terms of which the equations of motion and Bianchi identities read 
\be\label{eq:PCMeqmbianchi}
\begin{aligned}
\partial_+ {\mathcal J}_- + \partial_- {\mathcal J}_+ &=  0  \, ,  \\
\partial_+ {\mathcal J}_- - \partial_- {\mathcal J}_+ &=   [   {\mathcal J}_-,{\mathcal J}_+] \, .
\end{aligned} 
\ee
These currents are in fact the Noether currents for the $G_L$ symmetry; one could equally formulate the dynamics in terms of the currents of the $G_R$ symmetry.   These two equations can be repackaged in terms of a single $\frak{g}^\mathbb{C}$ valued connection \cite{Zakharov:1973pp}
\be
 {\mathcal L}_\pm(z) = \frac{1}{1\mp z}  {\mathcal J}_\pm \ , z \in \mathbb{C} \ . 
\ee
Here we have introduced an auxiliary complex constant $z$ that we call the spectral parameter.  This is useful since demanding that the Lax connection is flat i.e. that,
\be\label{eq:LaxFlat}
[\partial_+ + {\mathcal L}_+(z) ,\partial_- + {\mathcal L}_-(z)] = 0 \qquad \forall  z \in \mathbb{C} \, ,
\ee
for all values of the spectral parameter returns eqns.~\eqref{eq:PCMeqmbianchi}. What is gained?  In classical mechanical systems we are familiar with integrability in the Liouvillian sense, that there are sufficient involutive conserved charges (and corresponding symmetries) to reduce the dynamics to a foliation of level surfaces of the phase space on which action angle variables can be adopted to linearise the problem to a solvable set of ODEs.  In field theory  the continuum of degrees of freedom renders such a definition more subtle; however there is a sense in which there is an infinite number of symmetries.  These can be accessed via the monodromy\footnote{To avoid  complications that divert us from the main theme, we assume $\Sigma = \mathbb{R}^{1,1}$ and suitable asymptotics.} defined as 
\be
T(z) = P\exp\left[  -\int_{-\infty}^{\infty} \mathrm{d}\sigma\, {\cal L}_\sigma(z) \right]  \, . 
\ee   
By virtue of eq.~\eqref{eq:LaxFlat} this is conserved under time evolution and thus its expansion in $z$ leads to a tower of conserved charges,
\be
T(z) =  \exp \sum_{n}  z^{-(n+1)}  Q^{(n)} \, , 
\ee 
the first of which is nothing more than the global $G_L$ symmetry
\be
Q^{(0)}  = \int_{-\infty}^{\infty}   \mathrm{d}\sigma\,  {\mathcal J}_0  \, . 
\ee
The higher charges are non-local\footnote{There are also important higher-spin but local conserved charges \cite{Evans:1999mj} of which we shall not be discussing here.} and demonstrating the (Poisson) involutivity of these is a somewhat delicate point which we side step. The higher order charges give rise to an algebraic structure called a (classical) Yangian\footnote{The definition is not vital to us yet, though the interested reader is encouraged to look at the article of MacKay \cite{MacKay:2004tc} for a detailed treatment aimed at physicists.  Let us remark only that as explained by Drinfel'd \cite{Drinfeld:1987sy} the Yangian $Y(\frak{g})$ can be obtained as a limit of another algebraic structure, the affine quantum group $U_q(\widehat{\frak{g}})$ in which the quantum group parameter is sent to unity.  An obvious question is if  the quantum group structure can be realised as a symmetry of a sigma model.  We shall see later in these lectures this shall be achieved for any  $q \in \mathbb{R}$ with the so-called $\eta-$ or Yang-Baxter deformations of the PCM. The case of $q$  a root-of-unity, on the other hand, is realised by the $\lambda$-deformations of WZW models. 
 } $Y(\frak{g})$.  There is a useful freedom to perform a gauge transformation of the Lax and in particular one could chose to work instead with $  {\mathcal L}^g(z) = d g g^{-1} - g   {\mathcal L}(z)g^{-1}$. Performing the same construction will now result in a second Yangian's worth of charges, the first of which is that corresponding to the $G_R$ symmetry.  In this way what emerges is a rich infinite hidden symmetry of the type $Y(\frak{g}) \times Y(\frak{g})$.  
 
\begin{figure} 
 \includegraphics[width=\linewidth, clip, trim=1.5cm 11cm 4.5cm 6cm]{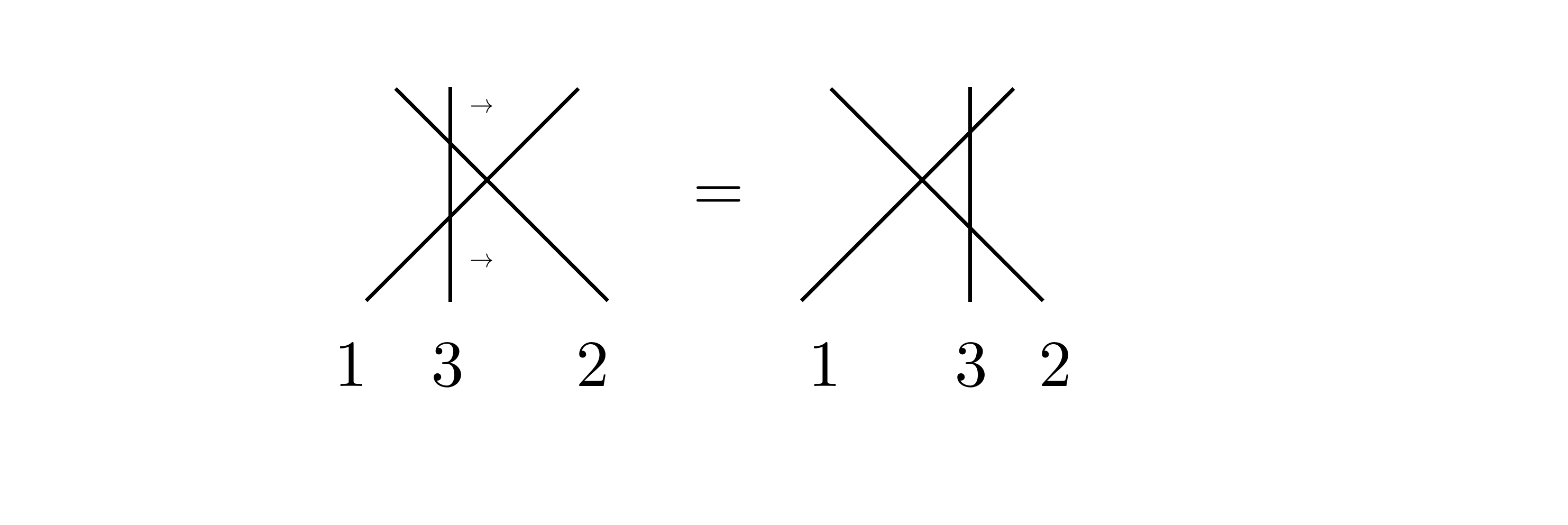} 
\vspace{-2cm}
\caption{\label{fig:tryagainYB}The factorised scattering giving the diagrammatic picture of the quantum Yang-Baxter equation}
\end{figure}

What of quantum integrability?  The essential idea of quantum integrability is that the symmetries, which are assumed to be unspoilt by quantum effects\footnote{See \cite{Goldschmidt:1980wq} for a discussion of anomalies that spoil quantum integrability.  In general when the target of the $\sigma$-model is a symmetric space, $G/H$, classical integrability is expected to survive in the quantum theory when $H$ is simple, but not otherwise.  As an example the bosonic theory on $\mathbb{CP}^n$ has an anomaly that spoils quantum integrability \cite{Abdalla:1980jt} which can however be restored though  the inclusion of appropriate fermions. }, constrain the dynamics to such an extent that the quantum S-matrix is completely determined by $2 \to 2$ body processes in which ingoing and outgoing momenta are simply permuted \cite{Zamolodchikov:1978xm}.     A $3 \to 3$ particle scattering can be factorised in terms of $2 \to 2$ processes and the result should not depend on the order of collisions  as depicted in figure \ref{fig:tryagainYB}.  
As a result the exact S-matrix $\mathbb{S}$, or rather the quantum R-matrix $\mathbb{R}$ of which it is built,  obeys the Quantum Yang-Baxter equation given schematically as \footnote{We suppress here spectral parameters, rapidities, and internal indices.} 
\be\label{eq:QYBE}
   \mathbb{R}_{13}  \mathbb{R}_{12} \mathbb{R}_{32}  = \mathbb{R}_{32}  \mathbb{R}_{12} \mathbb{R}_{13} \, . 
\ee

In this topic rather than momenta, massive states are labeled by rapidity $p_0 = m \cosh \theta$ and internal quantum numbers.  In the case of the $SU(N)$ PCM, for example, there are $N-1$ particles with masses $m_a = m \sin (\pi a /N)  / \sin(\pi / N)$, $a=1,2,\dots,N-1$, in multiplets transforming under the diagonal subgroup group of the $G_L \times G_R$  in the   $a^{th}$ antisymmetric tensor  representation.  The $2 \to 2$ S-matrix has a product form schematically given by
\be\label{eq:PCMSmatrix}
\mathbb{S}(\theta) =X(\theta)  S_{G_L} (\theta) \otimes S_{G_R}(\theta) \ ,  \quad \theta = \theta_1 - \theta_2 \, , 
\ee    
and in which $S_{G} (\theta)$ is $G$-invariant (in fact Yangian invariant) and $X(\theta)$ is an overall scale factor that has to be fixed by hand to ensure the correct analyticity properties of the S-matrix.  A rather beautiful result   that demonstrates the power of integrability is that an exact expression can be inferred \cite{Hasenfratz:1990zz,Balog:1992cm}, using thermodynamic Bethe Ansatze together with Wiener-Hopf  techniques, for the ratio  of the mass gap $m$ to the cut off $\Lambda$ (in the $\overline{M S}$  scheme, here for $G= SU(N)$):
\be
\frac{m}{\Lambda} = \sqrt{ \frac{8 \pi }{e} } \frac{N}{\pi} \sin(\pi  / N)  \, . 
\ee

 \subsection{Dualisation of the Principal Chiral Model} 

We now turn to the question of dualisation of the PCM.  As a first guess, motivated by the introduction, we might seek to exchange currents with Bianchi identities by introducing a set of dual scalar fields, $\Phi$, obeying
\begin{equation}
{\mathcal J}_\mu =  \epsilon_{\mu \nu}  \partial^\nu \Phi \, .
\end{equation}
The equations of motion for $\Phi$ follow from an action    
\be
S^\prime = - \kappa^2 \int \Tr \left(  \partial_\mu \Phi\partial^\mu \Phi   + \epsilon^{\mu \nu}[\Phi, \partial_\mu \Phi] \partial_\nu \Phi  \right) \, . 
\ee
However there are a few problems with this proposal \cite{Nappi:1979ig}.  Even classically though we have an on-shell map between degrees of freedom, the dynamics of this theory and the PCM are not truly related by a canonical transformation  \cite{Curtright:1994be}  as is the case with other duality transformations.  At the quantum level the coupling constant $\kappa$ in this {\em psuedo-dual} model has the opposite RG behaviour to that of the PCM.\footnote{In the lecture I presented some further folk law; that the {\em psuedo-dual} model exhibits particle production in its perturbative S-matrix \cite{Nappi:1979ig} and is thus unlike the PCM not quantum-integrable.  Recent work \cite{Hoare:2018jim} has shown this to be somewhat fallacious; the quantum-integrable PCM also exhibits particle production in $5$-point and $6$-point massless scattering possibly attributed to an IR regulator that breaks integrability.  One needs to treat arguments of perturbative particle production in massless scattering  with due caution in linking classical and quantum integrability. }  

We can do better if we adopt a path-integral dualisation {\em Buscher procedure} \cite{Buscher:1987qj}.  This is a three step recipe that starts with the PCM.
\begin{enumerate}
\item[Step 1:]  {\bf Gauge  the $G_L$ global symmetry}.  This is done by introducing a $\frak{g}$-valued connection, $A$, with which we can promote derivatives to covariant derivatives:
 $$ \partial g \mapsto Dg = \partial g + A g \, .$$
The PCM is then invariant under the local symmetry,  
\be
  g \to h^{-1}  g  \, , \quad A \to h^{-1} d h + h^{-1} A h \, .
\ee
 \item[Step 2:] {\bf Enforce a flat connection}.  So as to not change the number of physical degrees of freedom one should demand that this connection be pure gauge i.e. that the field strength, $F_{+-}=[D_+ , D_-]$, vanishes.  This is a done by appending to the gauged PCM action a Lagrange multiplier term  
\be
- \Tr v\cdot F_{+ - } =  v_a F^a_{+-}\, .
\ee
 Integrating out the Lagrange multipliers, $v_a$, essentially now returns one back to the starting PCM.  To do so we gauge fix $g=\mathbbm{1}$ and  then solve the constraint with 
 \begin{equation}
\label{eq:Aonshell} 
 A_\pm = \tilde{g}^{-1} \partial_\pm \tilde{g} \, .
\end{equation}
 After further redefining $\tilde{g} \to g$  original PCM is obtained.   
 \item[Step 3:]  {\bf Integrate out gauge fields}.  Here we again gauge fix $g=\mathbbm{1}$ and, after integrating by parts the derivatives in the Lagrange multiplier terms we solve for the non-propagating gauge fields 
   \begin{equation}
\label{eq:Aonshellv2} 
 A_+ = M^{-T} \partial_+v   \, , \quad   A_- = - M^{-1} \partial_- v \, .
\end{equation}
Here the matrix $M$, which comes from the quadratic term in gauge fields, is given by 
\be
M_{ab} = \kappa^2 \delta_{ab} - f_{ab}{}^c v_c \,, 
\ee
in which the portion proportional to $\delta_{ab}$ arises from the gauged kinetic term of the PCM and that proportional to the structure constants from the commutator contained inside the non-Abelian field strength.  Substituting these back into the action yields a T-dual sigma-model:
\begin{equation}\label{eq:Sdual}
\widehat{S}_{nAbT}[v] = \int \partial_+ v_a \left( M^{-1} \right)^{ab} \partial_- v_b \, .  
\end{equation}
\end{enumerate} 
Some comments are in order here and then a simple example to illustrate the technique.  

If we combine eqs.~\eqref{eq:Aonshell} and \eqref{eq:Aonshellv2} we have that 
\begin{equation}\label{eq:JtoV}
  {\mathcal J}_+ \mapsto M^{-T} \partial_+v \, , \quad  {\mathcal J}_- \mapsto - M^{-1} \partial_-v \ . 
\end{equation}
These equations, which involve  derivatives of fields, provide a  non-local  map between variables of the PCM and the T-dual.   Once cast in terms of canonical momenta and coordinates they constitute a canonical transformation\footnote{For the case of the $SU(2)$ PCM this relation was first established in \cite{Curtright:1994be}.}  of phase space variables \cite{Lozano:1995jx} with a generating function
\begin{equation}
F = \oint d\sigma~ \Theta(\sigma) \, , \quad   \Theta = v_a   {\mathcal J}^a_\sigma \, . 
\end{equation} 
Here $\Theta$ is a one-form pulled back to loop-space whose derivative $\omega = d\Theta = dv_a{\mathcal J}^a + \frac{1}{2} f_{bc}{}^a v_a {\mathcal J}^a\wedge {\mathcal J}^b  $ defines a symplectic form. 

Given this canonical transformation it is automatic that the T-dual theory is integrable and indeed the Lax connection is simply that of the PCM,   
\be
 {\mathcal L}_\pm(z) = \frac{1}{1\pm z}  {\mathcal J}_\pm \ ,\quad z \in \mathbb{C} \ , 
\ee
but with the currents now defined through the duality relation eq.~\eqref{eq:JtoV}. 

 The target space metric and B-field can be readily extracted from eq.~\eqref{eq:Sdual}.  Due to the appearance of the inverted coordinate dependent operator $M$,  these will typically be rather messy affairs that may not, in more general circumstances, exhibit any  isometry.  Any symmetry that does not commute with the dualised group $G$, including supersymmetry, will be lost (but recoverable non-locally on the world-sheet by virtue of the mapping eq.~\eqref{eq:JtoV}). 

{\bf Example:} As the simplest example let us consider $G= SU(2)$ such that the starting PCM describes the sigma model on $S^3$  equipped with the round metric.   There are then three Lagrange multiplier fields $v_a$, but the presentation of the dual is simpler if we transform them to spherical coordinates $\{r, \theta, \phi\}$.  The target space geometry is 
\be\begin{aligned}\label{eq:Tdual}
\widehat{ds}^2 &= \frac{dr^2}{\kappa^2} + \frac{r^2 \kappa^2}{r^2+\kappa^4}\left( d\theta^2 + \sin^2 \theta d\phi^2 \right) \, , \\
\widehat{B} &= \frac{r^3}{r^2 + \kappa^4}\sin \theta d\theta \wedge d\phi   \ . 
\end{aligned}
\ee
 This geometry looks like a fibration of a two-sphere over a line.  The $S^2$ indicates   a residual $G_R$ symmetry which commuted with the $G_L$ dualised.   In addition to these fields there is a non-constant dilaton profile generated by performing the Gaussian elimination of gauge fields in the Buscher procedure at the path integral level:
  \be
 \widehat{\Phi} = \phi_0 - \frac{1}{2}\log(r^2 + 
 \kappa^4 ) \, . 
\ee

{\bf Ornamentations:} Although we demonstrated the technique here with the simple dualisation of the group manifold, this approach can be easily extended with inclusion of spectator fields $Y^\mu$ to any target space for which we have a $G$ action acting without isotropy.  We take   the   general $\sigma$-model Lagrangian,
\be
L = E_{\mu \nu}(Y) \partial_+ Y^\mu \partial_- Y^\nu + E_{\mu a}(Y) \partial_+ Y^\mu L_-^a  + E_{ a \mu }(Y) \partial_- Y^\mu L_+^a +E_{ab}(Y) L_+^a L_-^b   \ , 
\ee
in which the $Y^\mu$ are inert spectators under a $G$ action, and in which the target space data in $E$ only depends on spectators.  The $L_\pm^i$ are the pull-backs of left invariant one-forms (essentially we are choosing adapted coordinates for the $G$ action).  An analogous Buscher procedure results in a T-dual model  
\be
\widehat{L} = E_{\mu \nu}(Y) \partial_+ Y^\mu \partial_- Y^\nu + \left( \partial_+ v_a + \partial_+ Y^\mu E_{\mu a} \right) (M^{-1} )^{ab}\left( \partial_- v_b +E_{b \mu  }  \partial_- Y^\mu \right)   \ , 
\ee
where $M_{ab} = E_{ab} - f_{ab}{}^c v_c   $. From this  Lagrangian  the T-dual target space data can be easily extracted. The canonical relations between T-dual models extends also to this inclusion of spectator fields \cite{Sfetsos:1996pm}.   

A slightly more involved situation is where $G$ acts with isotropy as is the case when the target space is a coset $G/H$. The first such example would be the three sphere, viewed as $SO(4)/SO(3)$ and equipped with the round metric, dualised with respect to the entire $SO(4) = SU(2)_L \times SU(2)_R$ isometry group. In doing a similar Buscher procedure one runs into the problem that there are more Lagrange multiplier fields ($\dim G$ of them) than the dimension of the  T-dual target space   (dimension $\dim G- \dim H$).  To handle this situation   requires a little care in the gauge fixing within the Buscher procedure.  Gauge fixing on the original coordinates, e.g. by assigning the coset representative parametrising $G/H$ some value, does not completely fix all the gauge symmetry. One is further required to also gauge fix $\dim H$ variables amongst the Lagrange multipliers; see  \cite{Lozano:2011kb} for some explicit treatments of this in a range of examples.  
   
 \subsection{Quantum considerations}
Since the seminal work of Rocek and Verlinde, \cite{Rocek:1991ps}, Abelian T-duality is firmly established as an exact quantum duality of CFTs   defined on a world sheet of any genus.  The situation with non-Abelian T-duality, and its generalisation of Poisson-Lie duality, is far less understood. 

The first quantum consideration is that of world sheet quantum corrections where  $\alpha^\prime$ is a loop counting parameter.  The picture that has emerged here is that the renormalisation of the starting theory at one-loop is matched in the dual theory.  That is to say if the starting theory is conformal, at least in the one-loop approximation where the metric is renormalised via Ricci flow plus contributions from the B-field, so too is the dual.  Equally if the starting theory is not conformal and the metric depends on some couplings that acquire a $\beta$-function, the T-dual metric will give rise to exactly the same set of  $\beta$-functions.\footnote{ There is a small caveat to this, when the dualised group is non-unimodular (i.e. $f_{ab}{}^{b} \neq 0$) then a Weyl anomaly will be ``produced'' by the dualisation \cite{Gasperini:1993nz}. To be more precise,  when the dualised group is non-unimodular the gauging of the Buscher procedure introduces a mixed gravitational-gauge anomaly \cite{Alvarez:1994np,Elitzur:1994ri} which is repackaged as a contribution to  a Weyl anomaly in the T-dual.  This artefact for non-semi-simple groups has more recently been reinterpreted,    \cite{Hoare:2016wsk,Hong:2018tlp}  in terms of producing backgrounds that solve the equations of  modified supergravity detailed in the appendix. }   We will see later that this one-loop compatibility of renormalisation can be made manifest by adopting a duality symmetric doubled formalism.   

  The question of what happens at higher orders in $\alpha^\prime$ is largely unchartered terrain, and even in the Abelian case is a rather involved affair. For instance, it is expected that Abelian T-duality will still act as a solution generating technique of the effective theory including $\alpha^\prime$/derivative corrections but the Buscher rules that determine the dual geometry also receive corrections \cite{Kaloper:1997ux}.  This topic is made even more challenging by the intrinsic ambiguities of scheme dependence and field redefinitions.   It seems plausible to me that the same will hold for non-Abelian duality but this is of course open.  

Let us turn to the issue of string genus $g_s$ effects.  One would like to employ the Buscher procedure to establish the dualisation of the theory in full string partition function given by the Polyakov sum over genus. A critical step of the Buscher procedure is to introduce a gauged $\sigma$-model together with constraints that ensure it is equivalent to the   starting theory.  Doing this carefully when the world sheet is a Riemann surface of genus $g$ requires keeping track of contributions from the holonomies of the gauge connection given by the path ordered exponential, $P\exp i \oint A$, of the periods around the $2g$ homology cycles.   In the case of Abelian T-duality, the path ordering becomes unimportant and demanding that the holonomies are the identity fixes the periodicity of the Lagrange multiplier (the T-dual coordinates).   However for the case of non-Abelian T-duality the winding modes of the Lagrange multiplier fields are insufficient to achieve this; first the Lagrange multipliers are adjoint valued and can not be assigned a natural periodicity in any case, and even if that were viable there is no obvious, or local, way to achieve path ordering of non-Abelian holonomies \cite{Giveon:1993ai}.  Of course it remains a possibility that some fix to $g_s$ corrections can be obtained (one can optimistically speculate that this will involve a full non-perturbative definition of the duality perhaps mediated by the introduction of D-branes or an open string sector) but the current understanding is that the non-Abelian ``duality'' transformation does not extend to the string genus perturbation theory. 

However, when considering holography at large $N$ and strong coupling we can largely suppress these interesting questions.  We are entitled to ask if non-Abelian T-duality can be usefully employed as a solution generating technique at the level of supergravity.  A precedent for this line of thinking comes from the work of Berkovits and Maldacena \cite{Berkovits:2008ic}. There it was shown that T-dualisation along effectively non-compact fermionic isometries of the Green-Schwarz superstring, which is afflicted with exactly the same quantum challenges, can be used to prove the Amplitude/Wilson loop connection of ${\cal N}= 4$ SYM at strong coupling.

 \subsection{Non-Abelian T-duality and Supergravity}

A critical question is then how to extend the above dualisation procedure to the full type II superstring including  the RR fields that enter type II supergravity.  The precise and correct approach should be to start with a relevant formulation of the superstring (Green-Schwarz or Pure-Spinor) but doing so can become rather burdensome with technology or calculation.   A more heuristic approach is to mimic closely the known results of Abelian T-duality, in particular the work of Hassan \cite{Hassan:1999bv}, and `bootstrap' the RR sector from the precise knowledge of the NS sector transformation rules \cite{Sfetsos:2010uq}.  That this procedure results in a supergravity solution was shown in numerous examples and a more rigorous proof was provided in \cite{Borsato:2017qsx}. 

The essential idea is to exploit the fact that after T-dualisation left and right movers couple to different frame fields for the same geometry.  Indeed, pushing forward  eq.~\eqref{eq:JtoV} to target space we have 
\begin{equation}
\widehat{ds^2} = \delta_{ab}\, e_L^a\otimes e_L^b  = \delta_{ab}\, e_R^a\otimes e_R^b \ , \quad e_L = - M^{-1} d v \, , \quad e_R = M^{-T} dv \, .
\end{equation}
This being the case, $e_L$ and $e_R$ must be related by a local Lorentz transformation 
\be
e_L = \Lambda\, e_R \ , \quad \Lambda = - M^{-1} M^T \, , \quad \Lambda^T = \Lambda^{-1} \, . 
\ee
Notice that $\det(\Lambda) = (-1)^d$ -- this indicates that the chirality of the theory will be changed for odd dimensional groups.  As is standard, this local Lorentz transformation induces an action on spinors via the Clifford map 
\be
\Omega^{-1} \Gamma^a  \Omega = \Lambda^a{}_b \Gamma^b \, . 
\ee
For the RR fields we work in the democratic formalism and consider the poly-forms i.e. the formal sum of forms of different degrees\footnote{The Hodge star acts such that  $\star^2 \omega_p = s (-1)^{p (D-p)} F_p$ in $D$ dimensions of signature $s$. }
\be
\begin{aligned}
IIA: ~~ \mathbb{F} &= F_0 + F_2 + F_4 - \star F_4 + \star F_2 - \star F_0 \ ,  \\ 
IIB:  ~~\mathbb{F} &= F_1 + F_3 + F_5 - \star F_3 + \star F_7   \ , 
\end{aligned}
\ee
which obey a Bianchi identity $(d-H\wedge)  \mathbb{F}= 0$.  
From these we can construct bi-spinors of $ Spin(d) \times  Spin(d)$  by contracting the components of differential forms with gamma-matrices i.e. $\slashed{F}_3= F^{(3)}_{abc}\Gamma^{abc}$ and so on. In terms of these objects the non-Abelian T-duality rule has a very compact form:
\be
e^{\widehat{\Phi}} \widehat{\slashed{\mathbb{F} }} = e^\Phi \slashed{\mathbb{F} }\cdot \Omega^{-1} \ . 
\ee
 The reason why $\Omega$ only attaches to one spinor index is roughly because its role is to translate a spinor index associated to a left-moving sector, and frame field, to a right-mover. Expanding out this formula and performing the appropriate gamma matrix manipulations we can extract the T-dual fluxes.  

Let's unpack this first to the case of a single $U(1)$ T-duality along a direction $\theta$ with $B=0$.  In that circumstance $\Lambda = - 1$, i.e. T-duality acts as parity on right movers, and the corresponding $\Omega= \Gamma^\theta\Gamma^{(11)}$. The T-dualisation rule now means that if we decompose $F^{(p)}= f^{(p)} +  g^{(p-1)} \wedge d\theta$ with $\iota_\theta f^{(p)} =0$, the transformation simply gives $\widehat{F}^{p+1} =     f^{(p)}\wedge d \tilde{\theta}  +  g^{(p+1)}$ and $\widehat{F}^{p-1} =     g^{(p-1)}  +  f^{(p-2)}\wedge d \tilde{\theta} $. 
 
We now return to our example based on the dualisation of the $G= SU(2)$ PCM to see schematically how this works in practice--and for full details the reader should consult the appendices of \cite{Itsios:2013wd}. In this case (we set $\kappa=1$ to suppress some irrelevant detail) one has that the Lorentz transformation is given by
\be
\Omega \sim  \frac{\Gamma^{123} + v_a \Gamma^a}{\sqrt{1+ v^2} } \, . 
\ee 
 The term with three gamma matrices means that a $p$-form $F^{(p)}$ can make contributions to $\widehat{F}^{(p\pm1)}$ {\em and} $\widehat{F}^{(p \pm 3)}$.   Now suppose  we embedded the PCM of this example inside $AdS_3 \times S^3 (\times T^4)$ supported in type IIB by pure RR three-form flux:
 $$
 F^{(3)} \sim \textrm{vol}(AdS_3) +  \textrm{vol}(S^3) \Rightarrow \slashed{ F }^{(3)} \sim \Gamma^{\tilde{0}\tilde{1}\tilde{2}} + \Gamma^{123} \, . 
 $$  
 When, for example, the $\Gamma^{123}$ in $\Omega$ is multiplied  against the same in  the flux, all gamma matrices annihilate leaving a zero-form, whilst when multiplied against the AdS part $ \Gamma^{\tilde{0}\tilde{1}\tilde{2}} $ a contribution to a six-form is obtained.   In this fashion we then find the T-dual is a solution of massive IIA  and even though we started with this simple background   the whole gamut of fluxes activated as schematically indicated in figure \ref{fig:fluxes}.

 \begin{figure}
\[
\begin{aligned}   \begin{tikzcd}
[%
    ,/tikz/column 1/.append style={anchor=base east}
    ,/tikz/column 2/.append style={anchor=base west}
     ,/tikz/column 3/.append style={anchor=base west}
          ,/tikz/column 4/.append style={anchor=base west}
    ]
      &  \widehat{F}^{(0)}    = &   \textcolor{blue}{1}    \\
  \textcolor{blue}{F^{(3)} } \arrow[rightarrow, blue, "\Gamma^{123}"]{ur}[swap]{}\arrow[rightarrow, blue, "v_a\Gamma^{a}"]{r}[swap]{ }  \arrow[rightarrow, blue]{dr}[swap]{ } \arrow[rightarrow, blue]{ddr}[swap]{ }\    &   \widehat{F}^{(2)}    = &   \textcolor{blue}{\frac{r^3}{1+r^2} \textrm{vol}(S^2) }  \\
     &  \widehat{F}^{(4)}      =&     \textcolor{blue}{r dr \wedge  \textrm{vol}(AdS_3)}  +  \textcolor{red}{\textrm{vol}(T^4)}   \\
    \textcolor{red}{ F^{(7)}} \arrow[rightarrow, red]{ddr}[swap]{ }\ \arrow[rightarrow, red]{ur}[swap]{ }\arrow[rightarrow, red]{r}[swap]{ }  \arrow[rightarrow, red]{dr}[swap]{ } 
 &  \widehat{F}^{(6)}    = & \frac{r^2}{1+r^2} \left(  \textcolor{red}{ r\textrm{vol}(T^4) \wedge\textrm{vol}(S^2)    } + \textcolor{blue}{ dr \wedge \textrm{vol}(AdS_3) \wedge\textrm{vol}(S^2)   }        \right)   \\ 
        & \widehat{F}^{(8)}  =&      \textcolor{red}{  -r dr \wedge \textrm{vol}(AdS_3) \wedge   \textrm{vol}(T^4)   } \\
          & \widehat{F}^{(10)}  =&     \textcolor{red}{ -  \frac{r^2}{1+r^2 } \textrm{vol}(AdS_3) \wedge dr \wedge \textrm{vol}(S^2)\wedge \textrm{vol}(T^4)   }  
   \end{tikzcd}
  \end{aligned}
  \]
 \caption{\label{fig:fluxes}Fluxes produced in dualisation of $AdS_3 \times S^3 \times T^4$.} 
 \end{figure}
 
 There is an alternative rather elegant way to state this dualisation as a Fourier Mukai transformation that makes more explicit the link to the canonical transformation induced by the T-duality.  Exactly parallel to the Abelian case \cite{Hori:1999me} one can consider $\omega = d\Theta $ with $\Theta = v_a {\cal J}^a$ as introduced previously as a field strength  and  use its Chern character as a kernel for a Fourier transformation \cite{Gevorgyan:2013xka}:
\be
 e^{\widehat{\Phi}} e^{-\widehat{B}}  \widehat{\slashed{\mathbb{F}}} = \int_G e^\Phi e^{-\omega}  e^{-B} \slashed{\mathbb{F} }  \, ,
\ee
in which the integration over $G$ should be understood as returning zero unless the integrand contains the the top-form with  $1= \int_G {\cal J}^1 \wedge \dots \wedge {\cal J}^d$.   Development of this approach may provide further insight to the topological or K-theoretic interpretation of non-Abelian T-duality.  

\subsubsection{$\star$ Supersymmetry $\star$  }  
{\bf Kosmann-Lie Derivative:}  A natural question to ask is what happens to (spacetime\footnote{See \cite{Sfetsos:1996xj} for a discussion of extended worldsheet supersymmetries under non-Abelian or Poisson-Lie T-duality.}) supersymmetry under this dualisation - how much of it is preserved?   The answer is quite elegant; Killing spinors of the original geometry will give rise again to Killing spinors in the dual, with a transformation rule rather similar to that of the RR fields, provided that they are invariant under the action of the dualised group.  Whilst the action of a vector on a generic spinor is not well defined, the action of a vector on a Killing spinor can be defined and is given by the spinorial Kosmann-Lie derivative \cite{Kosmann72}.   Letting $k_a$ be the vector fields that generate the $G_L$ action (i.e. the duals of the right invariant one-forms) a Killing spinor $\varepsilon$ is  invariant and preserved if   
\be\label{eq:LieKosmann} 
L_{k_a} \varepsilon =  k_a^i D_i \varepsilon - \frac{1}{4} \nabla_i k_{aj} \Gamma^{ij}  \varepsilon = 0   \, . 
\ee 
This idea was first used in the present context in \cite{Sfetsos:2010uq} where it was applied to  the  $AdS_3 \times S^3 (\times T^4)$ example described above.  In that example, exactly half, that can be isolated with a projector, of the sixteen super-symmetries are invariant and preserved in this sense.  A comprehensive analysis, \cite{Itsios:2012dc,Kelekci:2014ima}, shows indeed that when this condition holds the gravitino and dilatino supersymmetry variations vanish in the T-dual.  Moreover, when expressed in coordinates adapted to the $G$ action,  the condition eq.~\eqref{eq:LieKosmann}  states that the preserved super-symmetries are those for which  the corresponding Killing spinors do not depend on these coordinates. 

{\bf G-structures:} Many of the applications of non-Abelian T-duality have considered holographic duals to ${\cal N}=1$ supersymmetric gauge theory in four dimensions.  The conditions placed on the supergravity backgrounds in this context can be elegantly phrased in the language of  generalised geometry and in particular using G-structures \cite{Grana:2004bg}.  This is an interesting and extensive topic that we certainly can't do justice to in these notes, but the interplay of these ideas with non-Abelian T-duality \cite{Barranco:2013fza,Gaillard:2013vsa,Sfetsos:2014tza,Macpherson:2015tka} is worth outlining, albeit as a rough sketch. 

The central point is the characterisation of supersymmetries  in terms of bosonic objects built from bilinears of the two spinors that must exist on the the six-dimensional internal manifold $M$.          When the two spinors are parallel (i.e. proportional)  the structure group of $M$ is reduced to $SU(3)$   and we can construct a nowhere vanishing two-form $J$ and complex three form $\Omega_{hol}$   such that $J\wedge \Omega_{hol}= 0 $ and $J\wedge J\wedge J= \frac{3i}{4} \Omega_{hol}\wedge \bar{\Omega}_{hol}$.  From these we build ``pure spinors'' or poly-forms which,  suppressing some details of the normalisations, are given by
\be
\Psi_+ \propto  \exp - i J \ , \quad \Psi_-  \propto  \Omega_{hol} \, .
\ee
When the the two spinors are orthogonal then they each define  an $SU(3)$ structure, and the structure group on $M$ is reduced to the intersection of these, which is an   $SU(2)$.  Bosonically this implies a non-vanishing complex one-form $ v+ i w$, a real two-form $j$  and complex two-form $\omega$ packaged in to poly-forms 
 \be
\Psi_+ \propto  \exp( -i  v\wedge w) \wedge \omega   \ , \quad \Psi_-  \propto  \exp(- ij ) \wedge (v +  i w)    \, . 
\ee
More generally, the two spinors need not be orthogonal and the angle between them could   depend on position.   This more general case is called dynamic $SU(2)$ structure and and the expression for the pure spinors is modified.   In all cases, the conditions of supersymmetry (which together with Bianchi identities for the fluxes imply equations of motion) are very schematically given by 
\be
(d- H\wedge) \Psi_\mp  =  0 \ ,  \quad (d- H\wedge) \Psi_\pm  = F_{RR}    \ ,
\ee
in which $F_{RR}$ signifies the contributions from the RR fields and the choice of $\pm$ depends on IIA vs. IIB.

We now come to non-Abelian T-dualisation of these pure-spinors.   When asking if we should expect to preserve these objects under non-Abelian T-duality we can refer to  the Kosmann-Lie derivative condition of eq.~\eqref{eq:LieKosmann} which becomes the requirement that  $\Psi_\pm$    are invariant under the standard Lie derivative (extended to act on the formal sum of differential forms in the obvious fashion).   Under the T-duality $\Psi_\pm$  transform exactly as the RR fields do i.e. 
\begin{equation}
\widehat{\slashed{\Psi}}_\pm =   \slashed{\Psi}_\mp \cdot \Omega^{-1}  \ . 
\end{equation}  
Armed with this, together with the transformation of the RR fields, it is possible to give a rather direct proof that the non-Abelian duality is indeed a solution generating technique on a very wide class of interesting backgrounds.  A typical feature of the transformation is that the type of structure is changed, e.g. from $SU(3)$ to $SU(2)$.   Some examples of this are summarised in table \ref{tab:structurechange}.
\begin{table}
\begin{center}
  \begin{tabular}{ |l | c | r| }
   \hline
  	\textbf{Seed Solution} & \textbf{Seed Structure} & \textbf{Dual Structure}\\\hline
    \hline
    Klebanov-Witten & $SU(3)$ & Orthogonal $SU(2)$ \\ \hline
    Klebanov-Tseytlin & $SU(3)$ & Orthogonal $SU(2)$ \\ \hline
    $AdS_5\times Y^{p,q}  $ &  $SU(3)$  &Orthogonal $SU(2)$ \\ \hline
    Klebanov-Strassler & $SU(3)$ & Dynamical $SU(2)$  \\\hline
    KS Baryonic Branch & $SU(3)$ & Dynamical $SU(2)$  \\\hline
    Wrapped D5's on $S^2$ & $SU(3)$ & Dynamical $SU(2)$\\ \hline
    Wrapped D6's on $S^3$ & $SU(3)$ & Dynamical $SU(2)$\\ \hline
		Wrapped D5's on $S^3$ & $G_2$ & Dynamical $SU(3)$\\ \hline
    \hline
  \end{tabular}
\end{center}
 \caption{
  \label{tab:structurechange}Examples of the change of $G$-structures induced by non-Abelian T-duality with respect to an $SU(2)$ isometry group.}
\end{table} 

\subsection{Non-Abelian T-duality and Holography}

Armed now with a toolkit for creating new solutions of supergravity we seek to apply it in the context of holography.  The overarching motivation here would be to find supergravity solutions that `define' a strongly coupled (potentially even non-Lagrangian) quantum field theory,  and to carry out gravitational calculations that illuminate the properties of this hitherto less known postulated QFT dual. This strategy has been developed, with steadily increasing sophistication and elaboration to find novel supergravity solutions in a wide variety of contexts and dimensions by many authors.   The field theory interpretation of the geometries found in this manner has been developed across a number of the publications mentioned below but the most complete perspective is found in \cite{Lozano:2016kum,Lozano:2016wrs,Itsios:2017cew}.

The initial steps in this direction were taken in \cite{Sfetsos:2010uq} where the simplest examples of the dualisation of $AdS_3\times S^3$ and $AdS_5 \times S^5$ with respect to an $SU(2)$ non-Abelian isometry were conducted.  We will return to the case of $AdS_5 \times S^5$ viewed as a seed solution presently since it is rather prototypical of the general findings.  Before doing so let us briefly survey the, by now quite vast, zoology of solutions to supergravity constructed via non-Abelian T-duality.

For four-dimensional QFTs, the first exploration with less than maximal supersymmetry was taken in \cite{Itsios:2013wd} by considering the dualisation of the $AdS_5\times T^{1,1}$ Klebanov-Witten geometry, and later extended to more general $AdS_5\times Y^{p,q}$ solutions  \cite{Sfetsos:2014tza} which constitute examples of new families of ${\cal N}=1$ supersymmetric $AdS_5$ solutions in Type IIA and M-theory.  Beyond conformal field theories one can also   find supergravity solutions that encode rich RG dynamics of the QFT.  This   was achieved \cite{Itsios:2013wd,Kooner:2014cqa} by using the Klebanov-Strassler geometry as a seed solution to non-Abelian T-duality, or other RG flows like the Plich-Warner   \cite{Macpherson:2015tka,Dimov:2015rie}, and eventually paved the way for   construction of  backgrounds that geometrically encode confinement \cite{Gaillard:2013vsa} in terms of an internal manifold with a dynamic  $SU(2)$ structure.  

 $AdS_6$ backgrounds were found and interpreted in \cite{Lozano:2012au,Lozano:2013oma,Lozano:2018pcp}; in the context of $2+1$ dimensional QFTs $AdS_4$ solutions were constructed and analysed in \cite{Macpherson:2013zba,Lozano:2014ata,Zayas:2015azn,Lozano:2016wrs,PandoZayas:2017ier}.  Solutions with $AdS_3$ factors were studied in \cite{Lozano:2015bra,Bea:2015fja,Araujo:2015npa}.   Alongside these has been  the application  of this technique into diverse scenarioes: to plane waves and BMN limits  in  \cite{Itsios:2017nou,Lozano:2017ole};  to matrix models in \cite{vanGorsel:2017goj};  to non-relativistic QFT \cite{Araujo:2015dba} and  D-brane geometries   \cite{Terrisse:2018hhf}. 
 
\begin{figure}[h!]
 \begin{subfigure}{0.5\textwidth}
\includegraphics[width=0.9\linewidth, height=5cm]{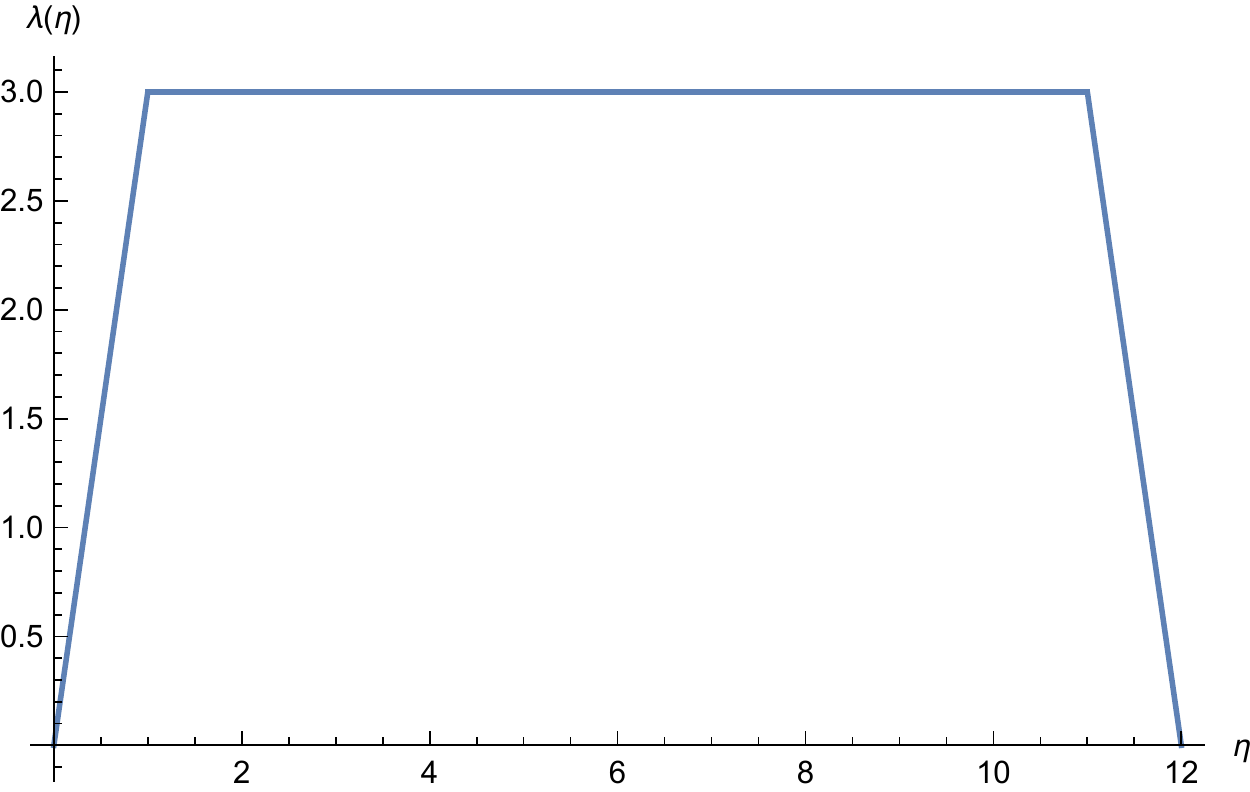} 
 \tikzstyle{gauge}=[circle,draw=blue!50,fill=blue!20,thick]
\tikzstyle{flav}=[rectangle,draw=black!50,fill=black!20,thick]
 \begin{center}
 \begin{adjustwidth*}{}{-0.10cm} 
 \begin{tikzpicture}[scale = 0.75] 
  \node (name1) at ( -1.25,0) [flav] {$N$};
    \node (name2) at (0,0) [gauge] {$N$};
    \draw[-] (name1.east) -- (name2.west);
        \node (name3) at (1.5,0) [gauge] {$N$};
            \draw[-] (name2.east) -- (name3.west);
          \node (name4) at (4,0) [gauge] {$N$};
        \draw[densely dotted] (name3.east) -- (name4.west);
          \node (name5) at (5.5,0) [gauge] {$N$};
                    \draw[-] (name4.east) -- (name5.west);
                           \node (name6) at (7,0) [flav] {$N$};    
                             \draw[-] (name5.east) -- (name6.west);
    \end{tikzpicture}
    \end{adjustwidth*}
    \end{center} 
\end{subfigure}
\begin{subfigure}{0.5\textwidth}
 \be
 \nonumber
  \lambda(\eta ) = \left\{ \begin{array}{cc} N \eta   &  0< \eta <1 \\ 
  N   &  1< \eta <K+1  \\ 
    N - N(\eta - K- 1)    &  K+1 <  \eta < K+2   \\ 
    \end{array} \right.  
  \ee
  \be
   \nonumber
   \begin{aligned} 
\sigma \partial_\sigma V   = \frac{N}{2} \sum_{n = -\infty}^{\infty} \sum_{l =1}^3 \sqrt{\sigma^2 + (\nu_l + n \Lambda - \eta)^2 }\\ 
\qquad\qquad  -  \sqrt{\sigma^2 + (\nu_l  - n \Lambda + \eta)^2 } \ , 
\end{aligned}
\ee 
 \phantomsubcaption 
\end{subfigure}
\caption{\label{fig:chargedensity} Charge density $\lambda(\eta)$ of  an ``Uluru'' spacetime (left top) with corresponding quiver (left bottom)  consisting of $K$ $SU(N)$ gauge group nodes terminated on each end by an $SU(N)$ flavour nodes ($N=3$, $K=10$ in plot).  The explicit form for solution  \cite{ReidEdwards:2010qs}  of $V$ given on the right, in which the parameters   $ \Lambda = 2 K+4$, $\nu_1 = 1$, $\nu_2 = K+1$, $\nu_3 = - K -2$.     }
\end{figure}
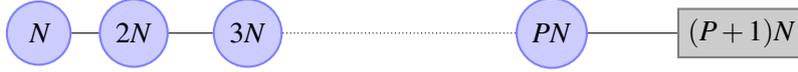

{\bf The dual of $AdS_5 \times S^5$:} Let us now focus attention to the example of the dualisation of $AdS_5 \times S^5$ with respect to an $SU(2)_L$ subgroup of the $SO(6)$ isometries of the five-sphere.  The $AdS_5$ part of the geometry is a  spectator to the dualisation, and we view the $S^5$   as a $S^2$ fibration over an $S^3$ on which the duality will act:
\be
ds^2(S^5) = 4 (d\theta^2 + \sin^2 \theta d\phi^2) + \cos^2 \theta ds^2(S^3) \, . 
\ee 
The result of the dualisation  follows directly from the treatment described previously\footnote{One essentially replaces $\kappa^2$, the radius of the PCM, with  $\cos^2 \theta$ in eq.~\eqref{eq:Tdual}.  } yielding a metric:
\be
\widehat{ds^2} = ds^2(AdS_5) + 4 (d\theta^2 + \sin^2 \theta d\phi^2) + \frac{dr^2}{\cos^2\theta} + \frac{r^2 \cos^2 \theta}{\cos^4\theta + r^2}  ds^2(S^2) \, .  
\ee 
This is completed with a dilaton, B-field and supported by RR $F_2$ and $F_4$ (together with their Hodge duals) as a solution of type IIA supergravity.   An analysis of the Kosmann-Lie derivative acting on the $AdS_5\times S^5$ Killing spinors, or a direct analysis of dilatino and gravitino equations, shows that the background is $\frac{1}{2}$-BPS and is a candidate dual to an ${\cal N}=2$ SCFT in four-dimensions.  Fortunately half-maximally supersymmetric backgrounds in IIA with $AdS_5$ factors are well understood through the work of Gaiotto and Maldacena \cite{Gaiotto:2009gz} as duals to classes of theories introduced by Gaiotto \cite{Gaiotto:2009we}.  In general the GM geometries are defined in eleven-dimensional supergravity and their construction requires solving a Toda-type equation in three dimensions.  However which the imposition of an additional $U(1)$ symmetry these can be reduced to solutions of IIA and the entire solution of supergravity can be mapped to the solution of a problem in electrostatics.  The whole geometry is described in terms of a potential function $V(\sigma, \eta)$ that solves the cylindrical Laplace equation,
\be
  \partial_{\sigma}[\sigma\partial_\sigma V]  + \sigma \partial_\eta^2 V = 0 \, ,
\ee
 such that the charge density located along the $\eta$-axis, $\lambda(\eta) = \sigma \partial_\sigma V |_{\sigma  =0}$, is piecewise linear and $\lambda(\eta= 0)=0$.  The dual gauge theories are linear quivers whose details can be  extracted  from the shape of the charge density profile.  In particular the ranks of the gauge group are given  from the value of $\lambda$ at integer arguments, and flavour groups are included at points at which the slope of $\lambda$ changes. For the quiver to be of finite length (and thus of finite central charge) requires that the profile $\lambda$ be a concave-down function with $\lambda$ restricted to an finite interval $\eta \in [0, \eta_\star]$ with  $\lambda(\eta= \eta_\star)=0$.    These cases were examined in \cite{ReidEdwards:2010qs}   and a typical example of this set-up is given in figure \ref{fig:chargedensity}.   One can relax the concavity requirement, at the expense of introducing some non-compactness in the geometry and having an unbounded quiver; this is what happens   in the case of the Maldacena-Nunez solution.  
 
 The T-dual of the $AdS_5\times S^5$ can be placed in this GM form with the identification 
 \be\label{eq:VST}
 \eta = \frac{r}{2} \ , \quad \sigma = \sin \theta \ , \quad V = \eta \log \sigma + \eta \left(\frac{\eta^2}{3} - \frac{\sigma^2}{2} \right) \ . 
 \ee
 We see immediately that the non-compactness of the $r$ direction is translated to a non-compactness in $\eta$, and since the charge density is linear given by $\lambda(\eta) = \eta$ this, taken literally, would suggest a quiver of infinite length with ever increasing ranks of gauge groups at its nodes: 
\tikzstyle{gauge}=[circle,draw=blue!50,fill=blue!20,thick]
\tikzstyle{flav}=[rectangle,draw=black!50,fill=black!20,thick]
\begin{center}  
 \begin{tikzpicture} \label{fig:onekinkquiver}
  \node (name1) at ( -1.25,0) [gauge] {$\,\,N\,$};
    \node (name2) at (0,0) [gauge]  {$2N$};
    \draw[-] (name1.east) -- (name2.west);
        \node (name3) at (1.5,0) [gauge]  {$3N$};
            \draw[-] (name2.east) -- (name3.west);
            \node (name4) at (3,0) [gauge] {$4N$};
    \draw[-] (name3.east) -- (name4.west);
       \node (name5) at (4.5,0) [gauge] {  $5 N$};
       \draw[-] (name4.east) -- (name5.west);
             \node (name6) at (6,0) {} ;
               \draw[densely dotted] (name5.east) -- (name6.west);
                              
    \end{tikzpicture}
\end{center}
A tricky point about the geometry above is that there is a singularity at $\sigma = 0$ that looks like an amalgamation of NS5 brane sources. 

The interpretation advocated in \cite{Lozano:2016kum,Lozano:2016wrs,Itsios:2017cew} is that the Buscher procedure we have used requires regularisation or completion so as to render the quiver finite. Whilst the geometry produced is acceptable for small values of $r$, in fact  $r$ should be restricted to lie in a finite interval.  Fortunately armed with the knowledge of the gauge theories we are given a natural way in which this should happen;  the quiver itself should be truncated and terminated with a flavour node: 
\begin{center}  
 \begin{tikzpicture} \label{fig:onekinkquiver}
  \node (name1) at ( -1.25,0) [gauge] {$\,\,N\,$};
    \node (name2) at (0,0) [gauge]  {$2N$};
    \draw[-] (name1.east) -- (name2.west);
        \node (name3) at (1.5,0) [gauge]  {$3N$};
            \draw[-] (name2.east) -- (name3.west);
    
       \node (name5) at (5.5,0) [gauge] {  $P N$};
               \draw[densely dotted] (name3.east) -- (name5.west);
                           \node (name6) at (8,0) [flav] {$(P+1) N$};    
                             \draw[-] (name5.east) -- (name6.west);
    \end{tikzpicture}
\end{center}
 This modification, whilst clearly ad-hoc, is a rather minimal way to make a holographic sense of the geometries found.  To achieve this modification one finds a potential  function $\dot{V}(\sigma,\eta)=\sigma\partial_\sigma V(\sigma,\eta)$   given by,
\begin{eqnarray}
& &  \dot{V} =  \frac{N}{2}  \sum_{n=-\infty}^{\infty} (P+1)\left[\sqrt{\sigma^2 +(\eta +P -2 n (1+P) )^2} - \sqrt{\sigma^2 +(\eta -P -2 n (1+P) )^2}   \right] +\nonumber\\
& & P\left[\sqrt{\sigma^2 +(\eta -1-P -2 n (1+P) )^2} -\sqrt{\sigma^2 +(\eta +1+P -2 n (1+P) )^2}  \right].\label{quiver1}
\end{eqnarray}
A virtue of this completion is the NS5-brane singularities at $\sigma =0$ have been pushed away to infinity giving a more calculable holographic space.  Since flavours are typically engineered via the inclusion of additional D-branes, this idea  hints at a non-perturbative (in $g_s$) completion to the non-Abelian T-duality transformation.  An amusing feature is that the string theory in the unbounded geometry described by the potential in eq.~\eqref{eq:VST} is integrable (at least classically in its bosonic sector) as a consequence of the canonical transformation corresponding to non-Abelian T-duality.  However, there is good evidence (numerical and analytic) that once the quiver is resolved with a flavour group as above  this geometry, or indeed the generic GM geometry, ceases to be integrable \cite{Nunez:2018qcj} (see also \cite{Wulff:2019tzh} for further recent remarks on non-integrability in GM geometries).  
 
 \section{The Doubled Worldsheet: Motivating the Drinfeld Double}

Double Field Theory \cite{Siegel:1993th,Siegel:1993xq,Hull:2009mi,Hohm:2010pp} and Exceptional Field Theory \cite{Berman:2010is,Hohm:2013vpa} have been key topics of the meeting here.  These are target space theories that seek to promote T- and U-dualities respectively to manifest symmetries.   In these lectures we are mostly working at the level of the worldsheet, and here too one can usefully find formulations in which duality is promoted to a manifest symmetry.  In the case of Abelian T-duality this doubled worldsheet has a long history  \cite{Duff:1989tf,Tseytlin:1990va,Tseytlin:1990nb,Hull:2004in,Hull:2006va} and is quite compelling; by promoting  string winding modes to appear on an equal footing with   momenta modes one can study T-fold string backgrounds in which T-dualities are used as a gluing transition of locally geometric spacetimes \cite{Hull:2006va}.  

In the case of non-Abelian T-duality, and indeed Poisson-Lie duality that we turn to shortly, there is also an equivalent doubled approach introduced in the seminal works of Klim\v{c}\'{i}k and \v{S}evera \cite{Klimcik:1995ux,Klimcik:1995ux,Klimcik:1996hp,Klimcik:1996nq} (see also \cite{Hull:2009sg} for a related proposal).  To present this let us return to the Buscher dualisation of the PCM on a group manifold $G$.  We will allow a   more general set up that includes a left-invariant (but not right invariant) metric and two form field on $G$ such that the non-linear sigma model becomes 
\be\label{eq:genPCM}
S= \int d^2\sigma~   L_+^a E_{ab} L_-^b \ , \quad E_{ab} = G_{ab} + B_{ab} \ , \quad g^{-1} \partial_\pm g =- i L_\pm^a T_a  \, , 
\ee
in which the entries of $E_{ab}$ are constant.  As before we gauge the $G_L$ symmetry by introducing gauge fields $A_\pm$  and introduce a Lagrange multiplier term $v_a F^{a}_{+-}$ that invokes a flat connection.  Instead of gauge fixing on $g$ and passing to the dual model, following \cite{Driezen:2016tnz} we  retain both fields $g$ and $v$ in a doubled formalism and perform  a {\em partial} gauge fixing\footnote{A Lorentz covariant gauge fixing is given by $A_\pm= \partial_\pm f$ and following this through leads to a Lorentz covariant Doubled approach with chirality constraints treated in the style of PST \cite{Pasti:1996vs}.} on $A_\pm$ in the spirit of \cite{Rocek:1997hi}:
\be
A_+ = A_- = \alpha \, .
\ee
The field $\alpha$ can be eliminated from the action and after  some rearrangement this yields
\be
S_{Doubled} = \int d^2\sigma~   -\mathbb{L}_\sigma^A  {\cal H}_{AB}  \mathbb{L}_\sigma^B   +   \mathbb{L}_\sigma^A  \eta_{AB}  \mathbb{L}_\tau^B + \mathbb{L}_\sigma^A  \Omega_{AB}  \mathbb{L}_\tau^B \, , 
\ee
in which the indices $A = 1 \dots 2 d$. Here the   matrices
\be\label{eq:HEtaOmega}
{\cal H}_{AB}  = \begin{pmatrix} G- B G^{-1} B & - BG^{-1}  \\ G^{-1} B & G^{-1} \end{pmatrix} \, , \quad \eta_{AB}  = \begin{pmatrix} 0  &  \mathbbm{1}_d \\  \mathbbm{1}_d    &   0  \end{pmatrix} \, , \quad \Omega_{AB}= \begin{pmatrix} 0  &  \mathbbm{1}_d \\  -\mathbbm{1}_d    &   0  \end{pmatrix} \, , 
\ee
are constant and define a para-hermitian or Born structure  (i.e.  define $I = {\cal H}^{-1} \Omega$ , $ J= {\cal H}^{-1} \eta$ and $K= \eta^{-1} \Omega$ with $-I^2 = J^2 =K^2 = \mathbbm{1}_{2d}$) which is a topic discussed in several talks in the meeting see \cite{Freidel:2013zga,Freidel:2018tkj}.  The field content, $g$ and $v_a$, has been packaged into a suggestive doublet
\be
\mathbb{L}^A(g,v)  = \begin{pmatrix} (g^{-1} d g)^a \\ D[g^{-1}]_{a}{}^b dv_b    \end{pmatrix} \, ,
\ee
in which we introduced the  adjoint action $D[g]_a{}^b T_{b}= g T_b g^{-1}$.  From this set up we have an action principle that gives rise to coupled first order equations for the fields $g$ and $v$ that, upon eliminating half the degrees of freedom in favour of the other half, return either the PCM equations of motion or that of its non-Abelian T-dual. 

One might be alarmed by the non-manifest Lorentz covariant nature of the action, and certainly this makes the study of the quantum behaviour of the double worldsheet more challenging.  Some reassurance should be given by the fact that in this derivation we began with a Lorentz invariant action  and it was only the adoption of a sort of axial gauge that caused this.

How should we interpret this result?   Suppose alongside the generator $T_a$ of $\frak{g}$ we introduce some generators for a second algebra $\tilde{\frak{g}} = \frak{u}(1)^d$ that we denote as $\tilde{T}^a$ and consider 
\begin{equation}
\tilde{g} = \exp  v_a \tilde{T}^a  \, . 
\end{equation} 
Formally then we see that\footnote{Thought of another way, we are picking a  local trivialisation of the cotangent bundle $T^\star G = G \ltimes \tilde{\frak{g}}$.}
\be
\mathbbm{g} = \tilde{g} \cdot g \ , 
\ee
can be used to construct 
\be
\mathbb{L}^A \mathbb{T}_A = \mathbbm{g}^{-1} d \mathbbm{g}  = (g^{-1} d g)^a T_a  + D[g^{-1}]_{a}{}^b dv_b     \tilde{T}^a  \, . 
\ee
Here we start to uncover an important structure, the (classical) Drinfeld double whose ingredients are: 
\begin{itemize}
\item A $2d$ dimensional Lie algebra $\frak{d}$ (group $\mathbb{D}$) with generators $\mathbb{T}_A$ and structure constants $\mathbb{F}_{AB}{}^C$
\item An adjoint-invariant symmetric inner product of  split  signature  $\eta_{AB} = \langle\langle \mathbb{T}_A, \mathbb{T}_B  \rangle\rangle$  such that $\mathbb{F}_{ABC}= \mathbb{F}_{AB}{}^D \eta_{DC}$ is totally antisymmetric
\item A decomposition into two $d$-dimensional sub-algebras $\frak{d} = \frak{g} \oplus \tilde{\frak{g}}$ which are both isotropic with respect to $\eta$. 
\end{itemize}
The last condition means that, if we pick a basis $\mathbb{T}_A = (T_a, \tilde{T}^a)$, we have $\eta_{AB}$ given exactly as  eq.~\eqref{eq:HEtaOmega}.  For the case at hand we have that  $\tilde{\frak{g}} = \frak{u}(1)^d$ but we should remain open minded to other possibilities for $\tilde{\frak{g}}$. However the choices of $\tilde{\frak{g}}$ that are compatible with a given $\frak{g}$ are restricted by   the Jacobi identity on the double $\frak{d}$.   
 
A final interesting feature is that
$$
d \left( \Omega_{AB} \mathbb{L}^A \wedge \mathbb{L}^B \right) = \mathbb{F}_{ABC}  \mathbb{L}^A \wedge \mathbb{L}^B  \wedge \mathbb{L}^C \, , 
$$  
which suggests viewing the doubled action as a chiral WZW model with $\partial{\cal M}_3 = \Sigma$:  
\be\label{eq:Sdoubled}
S_{{\mathbb{D}}}[\mathbbm{g}]  = \int_\Sigma d^2\sigma~   -\mathbb{L}_\sigma^A  {\cal H}_{AB}  \mathbb{L}_\sigma^B   +   \mathbb{L}_\sigma^A  \eta_{AB}  \mathbb{L}_\tau^B + \int_{{\cal M}_3}\mathbb{F}_{ABC}  \mathbb{L}^A \wedge \mathbb{L}^B  \wedge \mathbb{L}^C   \, . 
\ee
 In this  form we have an immediate generalisation to the case for which both $\frak{g} $ and $\tilde{\frak{g}}$ in the   double  $\frak{d} = \frak{g} \oplus \tilde{\frak{g}}$ are  non-Abelian subalgebras.  This generalisation is called Poisson-Lie duality and to it we shortly turn.
 
Before doing so, one nice thing about this doubled action is that we can calculate the one-loop beta functions that govern the renormalisation of the coupling constants contained inside $E_{ab}$  in a way that maintains the doubled covariance. The result is that  \cite{Avramis:2009xi,Sfetsos:2009vt}:
\be
\frac{d}{d \log \mu} {\cal H}_{AB} = \frac{1}{8} \left( {\cal H}_{AC}{\cal H}_{BF} - \eta_{AC}\eta_{BF} \right) \left( {\cal H}^{KD}{\cal H}^{HE} - \eta^{KD}\eta^{HE} \right) \mathbb{F}_{KH}{}^C \mathbb{F}_{DE}{}^F \, . 
\ee
The right-hand-side of this result is reminiscent of the equations of motion of Scherk-Schwarz reduced DFT, and indeed this is the case. Presented in this form the elegance of the result is rather obscured.  Instead we can work with the projectors viewed as elements of $\frak{d}\otimes \frak{d}^\star$
\be
{\cal P}_\pm = \frac{1}{2} \left(\mathbb{I} \pm {\cal H}^{-1} \eta \right)^A{}_B  \mathbb{T}_A\otimes \mathbb{T}^B \, , 
\ee
and a product on $\frak{d}\otimes \frak{d}^\star$ given by 
\be
[[ \cdot , \cdot ]] : (\mathbb{T}_A\otimes \mathbb{T}^B , \mathbb{T}_C\otimes \mathbb{T}^D ) \mapsto  [  \mathbb{T}_A,\mathbb{T}_C] \otimes  [  \mathbb{T}^B,\mathbb{T}^D] \, . 
\ee
 Then we have \cite{Klimcik:2018vhl} 
 \be 
  \mp \frac{d}{d \log \mu}    {\cal P}_\pm    =  {\cal P}_+ [[ {\cal P}_+,{\cal P}_- ]] {\cal P}_- + {\cal P}_- [[ {\cal P}_+,{\cal P}_- ]] {\cal P}_+ \, .
\ee
This later form makes explicit the appearance of projection operators that are needed to ensure compatibility of ${\cal P}_\pm ^2 ={\cal P}_\pm$ with renormalisation. 

\section{Poisson-Lie T-duality}

Let us now return to our general sigma-model eq.~\eqref{eq:sigmamodel} and consider the relaxation of the condition of isometry under a $G$-action that we suggested in the introduction.  Namely, for a $G$-action generated by vector fields $V_a = V_a{}^i \partial_i$ we have would-be-Noether currents 
\be
{\cal J}_{\pm~a} = V_{a}{}^i\left(G_{ij} \pm B_{ij} \right)\partial_\pm X^j\, ,
\ee 
which we demand obey a non-commutative modified conservation equation
\be
d\star {\cal J}_a  = \tilde{f}^{bc}{}_a \star  {\cal J}_b \wedge \star {\cal J}_c \, ,
\ee 
where $\tilde{f}^{bc}{}_a$ are some structure constants of an as yet unspecified algebra $\tilde{\frak{g}}$. 
In terms of the target space $E_{ij}= G_{ij} + B_{ij}$ this requires that
\be\label{eq:PL1}
L_{V_a} E_{ij} = \tilde{f}^{bc}{}_a V_b{}^m V_c{}^n E_{mi}E_{jn} \, ,
\ee
which is a rather non-trivial restriction and when satisfied the sigma model is said to possess a Poisson-Lie symmetry. The remarkable result of Klim\v{c}\'ik  and \v{S}evera \cite{Klimcik:1995ux} is that not only can these conditions be explicitly solved, but when they are solved there is a sense in which the sigma model can be T-dualised with respect to the $G$-action (even though it is not an isometry).  Before explaining how this is achieved, we take a short mathematical digression to  expose first the underlying algebraic structure  and   explain why the name Poisson-Lie is apt.  

Since the commutator of the Lie derivative of these vector fields should realise the algebra of $\frak{g}$, i.e. $[L_{V_a}, L_{V_b}]= f_{ab}{}^c L_{V_c}$ we find, after hitting eq.~\eqref{eq:PL1} with a second Lie derivative, a consistency condition 
\be\label{eq:cocycle}  
\tilde{f}^{ed}{}_a f_{bd}{}^c - \tilde{f}^{ed}{}_b f_{ad}{}^c 
+\tilde{f}^{cd}{}_b f_{ad}{}^e - \tilde{f}^{cd}{}_a f_{bd}{}^e  - \tilde{f}^{ec}{}_d f_{abv}{}^d= 0 \, .
\ee
This equation may look cumbersome but actually is well known in a different guise.  Let use consider the `dual' structure constants as defining a map 
\be \label{eq:cocycle2}
\begin{aligned}
\delta : \frak{g} &\to \frak{g}\wedge \frak{g} \, , \\
\delta(T_a) &\mapsto \tilde{f}^{bc}{}_a  \left( T_b\otimes  T_c  -T_c\otimes  T_b \right)  \, . \end{aligned} \ee
Then eq.~\eqref{eq:cocycle} has the interpretation that $\delta$ is a one-cocycle for $\frak{g}$ valued in  $\frak{g}\wedge \frak{g}$, i.e. in terms of an exterior derivative, $d$, acting in the Lie-algebra, $\delta$ is closed 
\be
0 = d \delta (x,y) \equiv \textrm{ad}_{x}\delta(y) - \textrm{ad}_y \delta(x) - \delta([x,y] ) \, , \quad \forall x,y \in\frak{g} \ . 
\ee
Moreover, that $\tilde{f}^{ab}{}_c$ obeys a Jacobi identity implies that $\delta$ obeys a co-Jacobi identity. The pair $(\frak{g}, \delta)$ is well known and called a Lie bi-algebra.    

A Poisson-Lie group is a group equipped with a Poisson bi-vector obeying a compatibility condition with the group multiplication rules. In general, a Lie group manifold inherits the algebra structure on its tangent space  $T_e G \cong \frak{g}$, and for a Poisson-Lie group the Poisson structure means that in addition $T^\star_e G = \frak{g}^\star$ is  equipped with a natural Lie-algebra structure compatible with that of $\frak{g}$.  The infinitesimal version of a Poisson-Lie group corresponds exactly to the Lie bi-algebra.  It is because of this connection that the name Poisson-Lie is used.

The Drinfeld double  introduced previously is essentially equivalent to this bi-algebra.  One considers $\mathbb{T}_A= (T_a, \tilde{T}^a)$ with $[T_a , T_b]= i f_{ab}{}^c T_c$ and $[\tilde{T}^a , \tilde{T}^b ] =i \tilde{f}^{ab}{}_c \tilde{T}^c$ then $\frak{d}= \textrm{span}_{\mathbb{R}}(\mathbb{T}_A)$ has a Lie algebra structure $[\mathbb{T}_A, \mathbb{T}_B]= i \mathbb{F}_{AB}	{}^C \mathbb{T}_C$ with mixed structure constants determined uniquely by demanding ad-invariance of $\eta_{AB} = \langle\langle \mathbb{T}_A, \mathbb{T}_B  \rangle\rangle$. The mixed components of the Jacobi identity for $\mathbb{F}_{AB}{}^C$ return again the cocycle constraint of eq.~\eqref{eq:cocycle}.

This is a very nice structure but do examples of such sigma-models really exist? Remarkably, yes!  We need to introduce a little further notation to demonstrate this.  Let $g \in G \subset \mathbb{D}$ ($\tilde{g} \in \tilde{G}$) and $L = g^{-1} dg$ ( $\tilde{L}= \tilde{g}^{-1} d \tilde{g}$) and define adjoint actions on $\frak{g}$ and $\tilde{\frak{g}}$ according to 
\be
g^{-1} T_a g = \mathbbm{a}_a{}^b T_b  \ , \quad g^{-1} \tilde{T}^a g = \mathbbm{b}^{ab} T_b + (\mathbbm{a}^{-1})_b{}^a \tilde{T}^b \ ,  
\ee
 and similar for the adjoint action of $\tilde{g}$. A most important combination is 
 \be\label{eq:pimat}
 \Pi^{ab}[g]  = \mathbbm{b}^{ca} \mathbbm{a}_{c}{}^{b}  = - \Pi^{ba}[g]\ , 
 \ee
 which is the Poisson bi-vector on $G$ obeying, for $h\in G$,  
\be
\Pi[g h ] = \mathbbm{a}^T[h] \Pi[g] \mathbbm{a}[h] + \Pi[h] \ , \quad \Pi[e]= 0 \ .
\ee
This group composition law is the analogue of the cocycle condition but now on $G$.  A  bi-vector on $\tilde{G}$, $\tilde{\Pi}[\tilde{g}]$, is defined analogously.

Now let $E_0 = G  + B $ be an invertible matrix with $d^2$ constant entries.  The Poisson-Lie sigma model on $G$ that obeys the condition  eq.~\eqref{eq:PL1} is given by 
\be\label{eq:PLact}
S_{G}[g] = \int L_+^a \left[ \left(E_0^{-1} + \Pi[g] \right)^{-1} \right]_{ab}L_-^b \, .  
\ee
Swapping the role of $g$ and $\tilde{g}$ gives a second sigma model  
 \be\label{eq:PLacttil}
\tilde{S}_{\tilde{G}}[\tilde{g}] = \int \tilde{L}_{+ a}  \left[ \left(E_0 + \tilde\Pi[\tilde{g}] \right)^{-1} \right]^{ab}\tilde{L}_{- b}   \, ,  
\ee
which again obeys the PL condition   eq.~\eqref{eq:PL1} but now with $f$ and $\tilde{f}$ interchanged.  Not only is it remarkable that the PL constraints can be solved, even more is that the two sigma-models $S$ and $\tilde{S}$ defined above are actually dual (at the minimum in the classical sense of being canonically equivalent  \cite{Klimcik:1995dy,Sfetsos:1996xj,Sfetsos:1997pi}).  This is especially remarkable given that the sigma model actions involve the inversion of rather complicated operators and correspond to potentially very involved  metric and B-fields.  To be explicit about the relation between $S$ and $\tilde{S}$ we define
canonical momenta  
\be  P_a = L^i{}_{a}\frac{\delta S}{\delta \dot{X}^i} \ , \quad L_\sigma^a = L^a{}_i \partial_\sigma X^i = (g^{-1} \partial_\sigma g)^a \, ,
\ee 
in which $ L^i{}_{a}L^a{}_j = \delta^i_j$  and likewise for tilde variables.  Then the canonical relation is given by 
\be
P = \tilde{\Pi} \tilde{P} + \tilde{L}_\sigma \ , \quad \tilde{P} = \Pi P + L_\sigma \, .
\ee

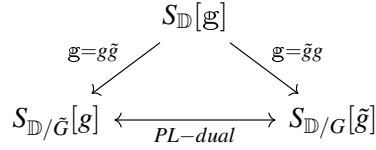
\begin{figure}
      \[   \begin{tikzcd}[column sep=small]
& S_{\mathbb{D}}[ \mathbbm{g} ]  \arrow[dl,  rightarrow]{}[swap]{\mathbbm{g}= g \tilde{g} } \arrow[dr,  rightarrow ]{}[]{\mathbbm{g}= \tilde{g} g } \\
  S_{\mathbb{D}/\tilde{G} }[g]    \arrow[leftrightarrow]{rr}[swap]{PL-dual} &                         &  S_{\mathbb{D}/ G }[\tilde{g}]
\end{tikzcd}\]
\caption{\label{fig:symplectoscheme}Relations between the doubled sigma-model and the two PL pairs.  Downward arrows indicate decomposition of group element in the double and subsequent integration out of degrees of freedom from the doubled action $ S_{\mathbb{D}}[ \mathbbm{g} ] $ defined in eq.~\eqref{eq:Sdoubled} .  The horizontal arrow indicates the canonical equivalence of Poisson-Lie dual pairs.  Here $\mathbb{D}/\tilde{G} \cong G$ and $\mathbb{D}/G \cong \tilde{G}$ (at least locally) and so $S_{\mathbb{D}/\tilde{G} }[g]$ and $S_{\mathbb{D}/G}[\tilde{g}]$  corresponds to the   actions  eq.~\eqref{eq:PLact}  and eq.~\eqref{eq:PLacttil} respectively.}
\end{figure}

Additionally, each of   $S$ and $\tilde{S}$ can be obtained from the doubled theory already introduced in the context of non-Abelian T-duality in \eqref{eq:Sdoubled} with now the Drinfeld double $\frak{d} = \frak{g} \oplus \frak{\tilde{g}}$ being fully non-Abelian as indicated in figure \ref{fig:symplectoscheme}. 

 \subsection{$\star$ A Target Space Perspective $\star$}  
Recent developments, some of which are reported on here at this workshop,  have   related these Poisson-Lie models to the target space DFT defined on the group manifold $\mathbb{D}$ \cite{Blumenhagen:2014gva,Hassler:2017yza,Demulder:2018lmj} and also \cite{Sakatani:2019jgu,Catal-Ozer:2019hxw},  elucidated the  Courant algebroids that underlie them \cite{Severa:2015hta,Jurco:2017gii,Severa:2018pag}, and developed the worldsheet doubled formalism \cite{Lust:2018jsx,Marotta:2018swj,Marotta:2019wfq}.   

For completeness, we briefly describe one of the most important features of the target space interpretation of the PL models.   We saw from the doubled world sheet that the $d^2$ {\em constant} parameters contained $E_0$ are naturally encoded in a ``flat'' generalised metric ${\cal H}_{AB}$.  The same is true for the curved target space data, $G_{ij}$ and $B_{ij}$, extracted from the non-linear sigma-model of eq.~\eqref{eq:PLact} which form a   ``curved'' generalised metric ${\cal H}_{\hat{I}\hat{J}}$.  Although $G_{ij}$ and $B_{ij}$ are in general rather complex affairs, when cast in terms of the generalised metric the situation becomes much simpler.  One can relate the flat and curved generalised metrics by\footnote{The superfluous hatting of objects here is to match conventions in \cite{Demulder:2018lmj}. } 
\be
{\cal H}_{\hat{I}\hat{J}} = \widehat{E}_{\hat{I} }{}^A  {\cal H}_{AB} \widehat{E}_{\hat{J} }{}^B \ , 
\ee
in which all the complexity is smuggled into the {\em generalised frame fields} $ \widehat{E}_{\hat{I} }{}^A $.  These objects, or more elegantly their inverses,   can be explicitly written in terms of $\Pi^{ab}$ and the components of the left invariant one-forms $L^a{}_i$ as 
\be
 \widehat{E}_A{}^{\hat{I} } = \begin{pmatrix} 1 & \Pi \\ 0 & 1 \end{pmatrix}\cdot  \begin{pmatrix} L & 0 \\ 0 & L^{-T} \end{pmatrix} \ . 
\ee 
This factorisation, into a $GL(d)$ action and $\beta$-transformation, makes it immediately apparent that the generalised frame fields are $O(d,d)$ elements.  One can view these $ \widehat{E}_{A}= (E_a, \tilde{E}^a)$ as a set of $d+d$ generalised vectors,  i.e. sections of the generalised tangent bundle  $TM+T^\star M$.  Without digressing too far into the topic of generalised geometry \cite{Hitchin:2004ut,Gualtieri:2003dx} we note for two such section $U= u+ \mu$ and $V= v + \nu$, there is a natural derivative, the Dorfman-derivative $\widehat{{\cal L}}$, that acts as 
\be
\widehat{{\cal L}}_U V= L_u v + \left( L_u \nu - \iota_\nu d \mu  \right)
\ee
in which $L_u$ is the conventional Lie-derivative and $\iota$ the interior product.  The remarkable hidden structure of the PL models becomes   apparent since the generalised frame fields exhibit an algebra
\be
\widehat{{\cal L}}_{ \widehat{E}_A}  \widehat{E}_B = \mathbb{F}_{AB}{}^C   \widehat{E}_C  \ . 
\ee

One can, of course, repeat this construction for the T-dual sigma-model of eq.~\eqref{eq:PLacttil} by using the analogue tilde quantities.  Then the T-duality transformation that relates the two models can be understood as \cite{Hassler:2017yza} performing coordinate dependent  $O(d,d)$  transformations to first undress the generalised frame fields on  ${\cal H}_{\hat{I}\hat{J}}$, then invert the flat space ${\cal H}_{AB}$ so as to implement $E_0 \mapsto E_0^{-1}$ and then redress with tilde generalised frame fields.    

A more in depth review of the ideas of this subsection, and their extensions, can be found elsewhere in the proceedings to this conference \cite{Demulder:2019bha}. 

 \subsection{Examples of Drinfeld Doubles}  
Let us close this section by giving a couple of examples of Drinfeld doubles which can be used to build PL models:
\begin{itemize}
\item $\frak{d} = \frak{u}(1)^d \oplus \frak{u}(1)^d$.  This is the case relevant to Abelian T-duality on a $d$-dimensional torus.  Here both $\Pi$ and $\tilde{\Pi}$ vanish, the one forms are simply $L^a = \partial \theta^a$ , $\tilde{L}_a = \partial \tilde{\theta}_a$ and the T-duality corresponds to the inversion $E = G+ B \mapsto \tilde{E}= (G+B)^{-1}$. 
\item $ \frak{d} = \frak{g} \oplus \frak{u}(1)^d$.  This is the case relevant to the non-Abelian T-duality.  Here $\Pi = 0$ and $\tilde \Pi = f_{ab}{}^c v_c$ with $\tilde{g}= \exp v_a \tilde{T}^a$ such that the $\tilde{S}$ is a generalisation of the non-Abelian T-dual eq.~\eqref{eq:Sdual} to the case where the PCM is equipped with a more general left invariant metric and B-field. 
\item Six-dimensional real Drinfeld doubles have been classified \cite{Snobl:2002kq}, and an example is $\frak{d}= \frak{so}(3,1)$. Closely related is the double  $\frak{d}= \frak{su}(2)^{\mathbb{C}} =  \frak{su}(2) + \frak{e}_3$ that is described in detail in the appendix and is useful for constructing explicit examples.
\item In general, the complexification   $\frak{d}= \frak{g}^{\mathbb{C}}$ viewed as a real algebra provides a Drinfeld double.  Under its Iwasawa decomposition $\frak{g}^{\mathbb{C}}   = \frak{g} + (\frak{a} + \frak{n})$ we have that $\frak{g}$ is the maximal compact subalgebra.  The ad-invariant inner product on $\frak{d}$ given by the imaginary part of the  Killing form on $\frak{g}^{\mathbb{C}}$.   At the level of groups $\mathbb{D}= G AN$ with $AN$ typically realised as some triangular matrices. 
\end{itemize} 
  
\section{Yang-Baxter and $\eta$-models}

The $\eta$ models that we shall describe below provide a very nice realisation of  PL dualisable models.  Not only that, they address a rather long standing question of what are the most general sigma models that are integrable. 

\subsection{Orientation: the squashed sphere} 
Let us return to the start; the PCM on the three-sphere.     It has been known since the work of Cherednik \cite{Cherednik:1981df} that this theory admits a current-current type deformation that preserves integrability 
\begin{equation}
\label{eq:PCMdef}
S_C[g]= -\frac{\kappa^2}{4\pi} \int {\textrm{d}}^2 \sigma\, {\textrm{Tr}}  \left( g^{-1} \partial_+ g g^{-1} \partial_- g \right) +  C J_+^3 J_-^3 \, , \quad J^3_\pm = i  {\textrm{Tr}}(g^{-1}\partial_\pm g T_3) \, .
\end{equation} 
This action defines a non-linear $\sigma$-model into a squashed three-sphere target space and whilst still integrable, the global symmetries are broken down to $SU(2)_L \times U(1)_R$.   What is rather amazing \cite{Kawaguchi:2011pf} is that the broken symmetry is recovered from  non-local charges\footnote{We review the construction of these charges in the appendix.} in the form of a semi-classical realisation of the quantum group $U_q(\frak{sl}_2)$ which begins with  
\begin{equation}
\label{eq:QG}
\{ Q^+_R , Q_R^- \}_{P.B.} =  i \frac{q^{Q^3_R} - q^{-Q^3_R} }{q - q^{-1}} \, , \quad q = \textrm{exp}\left(\frac{4\pi}{\kappa^2}\frac{\sqrt{C}}{1+ C} \right) \, .
\end{equation}
Deeper in  the tower of conserved charges extracted from the monodromy matrix one finds \cite{Kawaguchi:2012ve} additional (non-local) conserved charges   that enhance this to a semi-classical Poisson-bracket realisation  of a  quantum  affine algebra\footnote{The affine extension $\widehat{\frak{sl}_2}$ supplements  the Chevalley generators $\{E_1, F_1, H_1\}$ of $\frak{sl}_2$ with an additional root and corresponding generators $\{E_0, F_0, H_0\}$ obeying the standard relations $[H_i, E_j] = a_{ij} E_j$, $[H_i ,F_j] = -a_{ij} F_j$ and $[E_i, F_j]= \delta_{ij} H_j$ together with the Serre relations.  Here the generalised Cartan matrix $a_{ij}$ has off diagonal elements equal $-2$.   $K= H_0 + H_1$ is central and when  $K=0$, i.e.~for centreless representations,  then $\widehat{\frak{sl}_2}$ becomes the loop algebra.   This being the case, representations are the tensor of an $\frak{su}(2)$ representation and functions of a variable $z$.  There is a choice, known as gradation, to be made as to the relative action in  $\frak{su}(2)$  space and $z$-space.  In the  {\em{homogenous}} gradation   $E_1 = T^+ $, $F_1 = T^-$, $E_0 = z^{2} T^-$, $F_0 = z^{-2} T^+$, $H_1= - H_0 = T^3$.  In the  {\em{principal}} gradation $E_1 = z T^+ $, $F_1 = z^{-1}T^-$, $E_0 = z T^-$, $F_0 = z^{-1} T^+$, $H_1= - H_0 = T^3$.  These gradations lift to the quantum group deformation $  U_q(\widehat{\frak{sl}_2})$.  In the case described here, the  extension is centerless since $\tilde{Q}_R^3= - Q_R^3$ and the gradation is homogeneous. }   $U_q(\widehat{\frak{sl}_2})$.   The goal now is to generalise this spirit of integrability preserving deformations to the PCM based on any Lie group.

   \subsection{The modified Classical Yang Baxter equation } 
   
   A central role in the construction is played by the ${\cal R}$-matrix, an endomorphism of $\frak{g}$ that solves  the  modified classical Yang-Baxter equation (mCYBE)) 
\begin{equation}
\label{eq:mYB}
[{\cal R} x, {\cal R} y] - {\cal R}([{\cal R} x,y]+[x,{\cal R} y] ) = - c^2  [x, y]  \, ,  \quad \forall x,y \in \frak{g} \, . 
\end{equation}
Without loss of generality we can assume that $c^2 \in \{ -1,0, 1\} $  and the term ``modified'' refers to the possibility that $c \neq 0$ in the above (the case $c=0$ is referred to as CYBE).  We will assume that $ \cal R$ is skew with respect to the Killing form on $\frak{g}$ given as $\kappa_{ab} = \langle T_a , T_b \rangle$ so  that in components we write ${\cal R}(T_a) ={\cal R}_a{}^b T_b$ with ${\cal R}_a{}^c\kappa_{cb}= {\cal R}_{ab} =-{\cal R}_{ba}$ .     The mCYBE is an important equation, when solved it says that we can construct over the vector space $\frak{g}$ a second bracket 
\be
[x , y]_{\cal R} \equiv [{\cal R} x,y]+[x,{\cal R} y] \, , 
\ee
that obeys the Jacobi identity.  This equips  $\frak{g}$ (the vector space) with a second Lie algebra structure. To disambiguate things we call the vector space equipped with this bracket $\frak{g}_{{\cal R}}$.    We see the emergence of a bi-algebra structure that we will come back to.  Another way of thinking about the mCYBE is that it defines an algebra homomorphism 
\be
({\cal R}\pm c)  [ x , y]_{\cal R} = [({\cal R}\pm c) x,({\cal R}\pm c)  y] \, .
\ee
Sometimes it is natural to work with $\mathtt{r} \in \frak{g}\wedge \frak{g}$ given by 
\be
\mathtt{r}_{12} = \mathtt{r}^{ab}\frac{1}{2}\left(T_a\otimes T_b - T_b\otimes T_a \right) \, ,
\ee
where the subscripts indicate the location in a tensor product on which $\mathtt{r}$ acts (e.g. $ \mathtt{r}_{13} =   \mathtt{r}^{ab}T_a\otimes \mathbbm{1} \otimes T_b$ etc).  This is related to the above via
\be
{\cal R}(x) = \Tr_{2} ( \mathtt{r}_{12}\cdot \mathbbm{1}\otimes x )  \, , 
\ee
in which the trace acts on the second factor of the tensor product.  In a similar fashion one has the Casimir, $\mathtt{t}$, such that   $x=  \Tr_{2} ( \mathtt{t}_{12}\cdot \mathbbm{1}\otimes x ) $.   We can now define the operators $\mathtt{r}^\pm = \mathtt{r}\pm  c \mathtt{t}$  in terms of which the mCYBE takes an appealing form
\be\label{eq:CYBEv2}
0 = [  \mathtt{r}_{12}^\pm , \mathtt{r}_{13}^\pm  ] +  [  \mathtt{r}_{12}^\pm , \mathtt{r}_{23}^\pm  ] +  [  \mathtt{r}_{13}^\pm , \mathtt{r}_{23}^\pm  ]\, .
\ee

As an aside we comment that this classical YB equation has an origin from the quantum YB equation.  Recall the   quantum YB equation, which arises from considering factorised scattering,  constrains  a quantum  $\mathbb{R}$ -matrix to obey  
\be\label{eq:QYBE}
   \mathbb{R}_{13}  \mathbb{R}_{12} \mathbb{R}_{32}  = \mathbb{R}_{32}  \mathbb{R}_{12} \mathbb{R}_{13} \, . 
\ee
This equation  underpins much of quantum integrable systems. Let us assume that it admits a solution which can be expanded in some small parameter (conventionally denoted by $\hbar$)   
\be
  \mathbb{R}   = \mathbbm{1} + \hbar \mathtt{r} + O(\hbar^2)\, .
\ee
Then expanding out the quantum YBE eq.~\eqref{eq:QYBE} gives rise to the classical YB eq.~\eqref{eq:CYBEv2}.

In what follows we shall assume that ${\cal R}^3 = - {\cal R}$ (evidently weaker than ${\cal R}^{2}= - \mathbbm{1}$ allowing for the possibility that ${\cal R}$ has some zero eigenvalues).   For compact bosonic groups the  $c^2 = - 1$  solution is essentially unique e.g. for the case of $\frak{su}(2)$ one has ${\cal R}: ( T_1, T_2,T_3) \mapsto (- T_2 , T_1, 0)$. For  general semi-simple Lie algebra $\frak g$ this canonical ${\cal R}$-matrix   acts by anti-symmetrically swapping positive and negative roots and annihilates the Cartan.

 \subsection{The integrable $\eta$-deformed theory} 

Armed with such a solution to the mCYBE, Klimcik \cite{Klimcik:2002zj} proposed the following sigma-model
\be\label{eq:Seta}
S_{\eta}=- \frac{1}{t} \int d^2\sigma\, \langle  \partial_+g g^{-1} , \frac{1}{\mathbbm{1} - \eta {\cal R} } \partial_- g g^{-1} \rangle \, , 
\ee
in which we adopt the notation of angled brackets for the Killing form,  $\kappa_{ab} = \langle T_a , T_b\rangle$,  and have chosen to work in terms of right-invariant forms for reasons that will soon become apparent.  This is integrable \cite{Klimcik:2008eq}  and the Lax    (shown in \cite{Klimcik:2008eq} for $c^2=-1$ and for general  $c$ in \cite{Matsumoto:2015jja}) is given by:
\be
{\cal L}_\pm(z) = \frac{1}{1 \pm z} \left( 1 \mp \frac{\eta z}{1 \pm \eta {\cal R}} (-c^2 \eta \pm {\cal R} ) \right) \partial_\pm g g^{-1} \, . 
\ee
Let us first discuss the case of $c^{2}=-1$.  Here it was shown  by Delduc et al \cite{Delduc:2013fga} that, just as for the squashed $S^3$, there are  non-local conserved charges that give a classical version of a quantum group for the broken $G_L$ action.      Using techniques explained in appendix \ref{app:Maillet},  these charges can be identified by expanding the gauge transformed Lax  around special values of the spectral parameter, $z=   \pm i \eta$, that play a distinguished role in the Poisson Maillet algebra of the Lax.  For each simple root $\alpha_i$, of length $d_i = ( \alpha_i,\alpha_i)/2$, a   non-local conserved charge   $Q^{E_{\pm \alpha_i}}$ can be built whose Poisson-brackets   furnish   the algebra
\be
\{  Q^{E_{+ \alpha_i}},Q^{E_{-  \alpha_j}} \} = - i \delta_{ij}\frac{q^{d_i Q^{H_i}}  - q^{-d_i Q^{H_i}}     }{q^{d_i} - q^{-d_i} } \,  , \quad \{Q^{H_i} ,  Q^{H_j} \} = 0 \ , \quad \{  Q^{H_{ i}},Q^{E_{\pm  \alpha_j}} \} = \mp i A_{ij}  Q^{E_{\pm  \alpha_j} }\, , 
\ee
in which for a simple root $\alpha_i $ we have the distinguished Cartan generator $H_i$ such that $\kappa(H_i, H) =  \alpha_i(H) $ and $A_{ij}$ is the Cartan matrix.    Here the quantum group parameter is real, and taking into account the overall normalisation of the sigma-model, is given by $q= \exp( 4 t\eta)$ which is an RG invariant (at one loop at least).  In addition to the above relations, there is also a Poisson bracket implementation of the q-Poisson-Serre relations. What was also made clear in \cite{Delduc:2013fga} was that the above algebra can be obtained as a ``classical'' limit $\hbar \to 0 $ of (the compact real form of) the quantum group $U_{\tilde{q}}(\frak{g}^{\mathbb{C}})$ with $\tilde{q} = q^{\hbar}$ under  the replacement of commutators $\hbar^{-1} [ \cdot, \cdot] \to i \{ \cdot, \cdot \}_{P.B.}$.     
As with the case of the $\sigma$-model on the squashed three-sphere \cite{Kawaguchi:2012ve}, this structure is supplemented \cite{Delduc:2017brb} with additional  conserved charges, located deeper in the expansion of the monodromy,  that upgrade this structure to a quantum affine algebra $U_q(\widehat{\frak{g}})$.   The symmetry that generates these charges has an action on fields encodes in a non-Abelian moment map as detailed in \cite{Delduc:2016ihq}.

 \subsection{The $\eta$-deformed theory as a PL model } 

Let us do a little manipulation that will allow us to relate the $\eta$-model to a PL sigma-model:
\be
\begin{aligned} 
S_{\eta}& =- \frac{1}{t} \int d^2\sigma~   \langle \partial_+g g^{-1},  \frac{1}{\mathbbm{1} - \eta {\cal R} } \partial_- g g^{-1} \rangle\\
& =- \frac{1}{t\eta} \int d^2\sigma~   \langle  L_+,   (\textrm{ad}_{g})^{-1}  \left(\eta^{-1} \mathbbm{1} -  {\cal R} \right)^{-1} (\textrm{ad}_{g^{-1}})^{-1} L_- \rangle\\
& =- \frac{1}{t\eta} \int d^2\sigma~  \langle  L_+,    \left(\eta^{-1} \mathbbm{1} - \textrm{ad}_{g^{-1}} {\cal R}\textrm{ad}_{g}  \right)^{-1}    L_- \rangle  \, ,  \end{aligned}
\ee
in which $L_\pm = g^{-1} \partial_\pm g$.  This last form is indeed that corresponding to the Poisson-Lie sigma model eq.~\eqref{eq:PLact} with the identification 
\be\label{eq:EPi}
E_0^{-1} = \eta^{-1} \mathbbm{1} - {\cal R} \ , \quad \Pi[g] = \textrm{ad}_{g^{-1}} \cdot {\cal R} \cdot \textrm{ad}_{g}  -  {\cal R} \, , 
\ee
in which indices are raised appropriately with $\kappa_{ab} = \langle T_a , T_b\rangle$.
To understand this result at an algebraic level we recall that a one-cocycle, $\delta$, valued in $\frak{g}\wedge \frak{g}$ is said to be co-boundary if 
\be
\delta(x)=  [ x, \mathtt{r} ] \ , \quad \mathtt{r} \in  \frak{g}\wedge \frak{g} \, , 
\ee  
which in component form requires that (c.f. eq.~\eqref{eq:cocycle2} ) 
\be
\tilde{f}^{ab}{}_c  = \mathtt{r}^{ae} f_{ce}{}^b -  \mathtt{r}^{be} f_{ce}{}^a \ , 
\ee
which one immediately recognises as the structure constants of the Lie-algebra $\frak{g}_{{\cal R}}$.  Notice further that if we set $g = \exp( \epsilon x)$  we have that 
$\Pi[g]  =  \epsilon [ x, \mathtt{r} ]   + O(\epsilon^2)   =   \epsilon \delta(x)  + O(\epsilon^2)$, which has the interpretation  that $\Pi$ is a cocycle on $G$ obtained as the integral of the co-boundary cocycle on $\frak{g}$.   Together $\frak{g}$ and $\frak{g}_{\cal R}$ define a   bi-algebra  and hence a double  $\frak{d} = \frak{g} \oplus \frak{g}_{\cal R}$. 

For the case of $c^2 = - 1$ the canonical ${\cal R}$-matrix associated to a semi-simple Lie algebra $\frak g$ with Killing form $\kappa = \langle \cdot , \cdot \rangle$    acts by anti-symmetrically swapping positive and negative roots and annihilates the Cartan.  In this case  the double is the direct sum $\frak d= \frak g\oplus \frak g_{\cal R}$, which is isomorphic to the complexification $\mathfrak g^{\mathbb C}$ of $\frak g$. This double can be decomposed using the Iwasawa decomposition $\frak g^{\mathbb C}=\frak g \oplus (\frak a \frak n)$, where   $\frak g$ and $\frak a \frak n$ are both isotropic subalgebras of $\frak d = \frak g^{\mathbb C}$. The ad-invariant non-degenerate symmetric bilinear form on $\frak d = \frak g^{\mathbb C}$ is 
\begin{align*}
\langle \langle Z_1,Z_2\rangle \rangle = -i\, \langle Z_1,Z_2 \rangle + i\,\overline{\langle Z_1,Z_2\rangle} \,,
\end{align*}
in which   $Z_i \in \frak{g}^{\mathbb C}$ and $\overline{\cdot}$ is complex conjugation.
To tie this discussion to the doubled sigma model, following \cite{Klimcik:2015gba}, we now  simply need to identify the generalised metric, which follows from eq.~\eqref{eq:EPi} as 
\begin{equation}
{\cal H}_{AB} = \begin{pmatrix} \eta \kappa_{ab}  & - \eta {\cal R}_{a}{}^b \\  \eta {\cal R}^{a}{}_b & \frac{\kappa^{ab}}{\eta} - \eta {\cal R}^{a}{}_c {\cal R}^{c b}  \end{pmatrix} \, , 
\end{equation} 
in which indices on ${\cal R}$ are raised with $\kappa$.  Further perspective of how these models are related to the doubled approach are found in \cite{Lust:2018jsx,Demulder:2018lmj}. 

We see then that $\eta$-deformed PCM can be cast as a PL $\sigma$-model and hence admits a dual description.
 We will return to discuss the significance of this T-dual theory at the very end of the report. 

 \subsection{The  $c=0$ case,   TsT and non-Abelian T-duality } 
 
 For compact bosonic algebras there is no solution of the mCYBE with $c^2>0$; we will not discuss this case further.  The $c=0$ case, i.e. that built on a solution of the non-modified CYBE,  holds interest since  \cite{Matsumoto:2014nra,Osten:2016dvf}  it very nicely encodes things like TsT transformations  that  generate holographic backgrounds corresponding to non-commutative $\beta$-deformations \cite{Lunin:2005jy}.  
 
This $c=0$ case can also be understood in terms of non-Abelian T-duality \cite{Hoare:2016wsk,Borsato:2016pas,Borsato:2018idb}.  Consider again the T-dualisation of the PCM on a group $G$ with respect to a subgroup $H_L$ (with generators $T_\alpha$) of the $G_L$ isometry,    but now following \cite{Borsato:2016pas} include a  pure-gauge B-field contribution 
\begin{equation}
B = \beta \omega_{ij} dx^i\wedge dx^j = \zeta  \omega_{\alpha \beta } (g^{-1}dg)^\alpha  \wedge  (g^{-1}dg)^\beta \, . 
\end{equation} 
Of $\omega$ we require that $H=dB = 0 $ and that the  $H_L$ action is not spoilt.  Together these imply that $\omega$ defines a two-cocycle.   Performing the $H_L$ (non-Abelian) T-dualisation results in a dual model.  If $H_L$ is Abelian this indeed gives rise to a TsT transformation.  Note that   $B$ is   pure gauge when $\omega$ is co-boundary,  $\omega(x,y) = f( [x,y ])$, so non-trivial deformations are labelled by elements of the cohomology $H^2(\frak{h}_L)$.  

A second way to achieve the same result is to first centrally extend the dualised isometry group $\frak{h}_L$ and then T-dualise.  The Lagrange multiplier field corresponding to central directions in the algebra turn out to be non-dynamical and frozen to a constant value \cite{Hoare:2016wsk}. One can harness all the tricks of non-Abelian duality of the first section, e.g. recovering all the RR fields in the duals to marginal/non-commutative deformations of ${\cal N}=4$ SYM \cite{Hoare:2016wca}.  The  approach \cite{Hoare:2016wsk} is equivalent to that described above \cite{Borsato:2016pas} essentially because the central extension 
\be
[T_a, T_b]= f_{ab}{}^c T_c + \omega_{ab} Z  \ , \quad [T_a, Z]= 0  \, ,
\ee       
 satisfies the Bianchi identity when $\omega$ obeys the cocycle condition.   
 
 How then do these relate back to the YB models?  The idea is that when the CYBE holds, $\mathtt{r}^{ab}$ is invertible on a sub-algebra and its inverse defines exactly the cocycle $\omega$ introduced above.  To complete the matching of  the YB model eq.~\eqref{eq:Seta} with this dualisation plus pure-gauge B-field  requires a) that deformation parameters are matched $\beta \sim \eta^{-1}$ and b) a non-trivial field redefinition whose details are found in \cite{Borsato:2016pas}.

 \subsection{ $\star$ Multi-parameter  $\eta$-deformations  $\star$} 
 
 In the case of $\eta$-deformations on group manifolds there are  elegant extensions that introduce additional deformation parameters but still preserve integrability.  

 The first of these is the bi-Yang-Baxter sigma-model  \cite{Klimcik:2014bta,Klimcik:2016rov} with the action 
 \be 
S_{\eta,\zeta}  =- \frac{1}{t} \int  d^2 \sigma~  \langle \partial_+g g^{-1},  \frac{1}{\mathbbm{1} - \eta {\cal R}  - \zeta {\cal R}^g  } \partial_- g g^{-1} \rangle \ ,
\ee
in which the explicit field dependence in the operator $ {\cal R}^g =  \textrm{ad}_{g^{-1}} \cdot {\cal R} \cdot \textrm{ad}_{g}$ means that both left and right global symmetries are broken by the deformation (though both are restored in the non-local quantum group sense). 
 
The second of these  \cite{Delduc:2014uaa,Klimcik:2017ken,Demulder:2017zhz}, extending the initial work \cite{Kawaguchi:2011mz,Kawaguchi:2013gma} done for $SU(2)$,  is to add a Wess-Zumino term\footnote{The next  section on $\lambda$-deformations gives more details of the WZ coupling and WZW model.} to the PCM   with the action
 \be 
S_{\eta,\zeta}  = - \frac{1}{2 \pi } \int  d^2 \sigma~   \langle \partial_+g g^{-1} \left( \alpha \mathbbm{1} +\beta  {\cal R}  +\gamma {\cal R}^2 \right) \partial_- g g^{-1} \rangle  + \frac{k}{24 \pi } \int_{M_3} \langle \bar{g}^{-1} d \bar{g} , [ \bar{g}^{-1} d \bar{g},\bar{g}^{-1} d \bar{g}] \rangle \ .
\ee
 This is integrable on a locus of coupling constants \cite{Delduc:2014uaa}
 \be
 \beta^2 = \frac{\gamma}{\alpha} \left( \alpha^2 - \alpha \gamma - k^2 \right) \, ,  
 \ee  
a condition that is preserved consistently along the RG flow \cite{Demulder:2017zhz}. In addition to $k$, there is another RG invariant $\Theta^2 = \alpha^2 \beta^2 \gamma^{-2}$. For the case of $SU(2)$ it was shown  \cite{Kawaguchi:2013gma} that the deformed right acting symmetry is  similar\footnote{It is not precisely $U_q(\widehat{\frak{su(2)}})$ since the  relations that move up and down the gradation are modified.} to an affine quantum group $U_q(\widehat{\frak{su(2)}})$ with parameter determined by the RG invariant
$$
q= \exp\left[ \frac{2\pi\Theta }{\Theta^2 + k^2 } \right] \, . 
$$  
This model holds additional appeal since it is self-dual under Poisson-Lie T-duality \cite{Klimcik:2017ken,Demulder:2017zhz} with a simple map inversion map $\alpha \to \frac{k^2}{\alpha}$ reminiscent of Abelian T-duality.     Indeed, the self dual point benefits from enhanced symmetries; it is the conformal  WZW model realised at the IR fixed point of the RG flow.  

Further, both these deformations can be combined together, along with ones of the class $c=0$, and on a suitable locus of coupling parameters they give an integrable model whose Lax was explicitly realised in \cite{Delduc:2017fib}.

 \subsection{  $\star$ $\eta$-deformations on symmetric and semi-symmetric spaces $\star$ } 

Let us briefly mention the case of bosonic symmetric spaces and  super-cosets relevant to e.g. the type II string on $AdS_5\times S^5$. 

We first consider a $\mathbb{Z}_2$ graded bosonic algebra $\frak{g} = \frak{g}^{(0)} + \frak{g}^{(1)}$ with $\frak{g}^{(0)}\equiv \frak{h}$ a sub-algebra (the fixed set of an involution of order 2) with corresponding group $H$, and $[ \frak{g}^{(i)},\frak{g}^{(j)}]\subset \frak{g}^{(i+j~\textrm{mod}~2)}$.  Then $G/H$ is a symmetric space and the sigma-model on this space is integrable and can be written in terms of an action 
\be
S = -\frac{1}{t} \int d^2\sigma ~ \langle  L_+,  P^{(1)}(L_- )  \rangle\ , \quad 
\ee
in which $P^{(i)}$ is a projector onto $\frak{g}^{(i)}$ and $L_\pm = g^{-1} \partial_\pm g$.  Notice   under local $H$ right actions, $g\to g h  $,   the coset projection of the left-invariant one-forms transform covariantly $L_\pm^{(1)} \to h^{-1} L_\pm ^{(1)} h = \textrm{ad}_{h^{-1}} L_\pm ^{(1)}  $. Given the ad-invariance of the inner product,  the action is invariant under this local symmetry and consequently  one can fix away $\dim(H)$ coordinates  by  choosing the group element to be a coset representative.    This theory is again integrable (in the classical Lax sense at least) with a connection 
 \be
 {\mathcal L}_\pm(z) = L^{(0)}_\pm + z^{\pm 1} L^{(1)}_\pm \ , ~~~ z \in \mathbb{C} \ . 
\ee
The $\eta$-deformed theory \cite{Delduc:2013fga} is again constructed with a solution of the mCYBE (here we restrict to $c^{2}=-1$) with action   
\be
S_{\eta}[g] = -\frac{1}{t}  \int d^2 \sigma\,  \langle  L_+, \frac{1}{\mathbbm{1} - \eta {\cal R}^g \cdot P^{(1)}  }  P^{(1)}(L_- )   \rangle\ , \quad 
\ee
in which $ {\cal R}^g =  \textrm{ad}_{g^{-1}} \cdot {\cal R} \cdot \textrm{ad}_{g}$ has been dressed to ensure that under the local  $H$ right action   ${\cal R}^g  \to \textrm{ad}_{h^{-1}} \cdot  {\cal R}^g  \cdot \textrm{ad}_{h}$ and that the gauge invariance is preserved.  To define a Lax for this theory we introduce the currents 
\be
A_\pm = B^{(0)}_\pm + \sqrt{1+\eta^2} B^{(1)}_\pm \quad \textrm{with} ~~~~~    B_\pm = (\mathbbm{1} \pm  \eta {\cal R}^g \cdot P^{(1)})^{-1}  L_\pm \ . 
\ee
Then the Lax \cite{Delduc:2013fga} has the same form as the undeformed theory written in terms of these modified currents 
 \be
 {\mathcal L}_\pm(z) = A^{(0)}_\pm + z^{\pm 1} A^{(1)}_\pm \ , ~~~ z \in \mathbb{C} \ . 
\ee

To include fermions we work on a $\mathbb{Z}_4$ graded super-algebra $\frak{g} = \frak{g}^{(0)} + \frak{g}^{(1)} + \frak{g}^{(2)}+ \frak{g}^{(3)}$  with $\frak{g}^{(0)}= \frak{h}$ a bosonic sub-algebra, $\frak{g}^{(2)}$ bosonic, and with $\frak{g}^{(1)}$ and $\frak{g}^{(3)}$ fermionic. We introduce also a projector
\be
\Omega_\eta = P^{(1)} + \frac{2}{1-\eta^2} P^{(2)}- P^{(3)} \, . 
\ee
The undeformed Metsaev-Tseytlin \cite{Metsaev:1998it} action is given by 
\be
S = -\frac{1}{2 t}  \int d^2\sigma (\eta^{\mu\nu} - \epsilon^{\mu \nu}  )\langle L_\mu  , \Omega_0 \cdot L_\nu   \rangle\ \sim \int \textrm{STr}( L^{(2)}\wedge \star  L^{(2)} -  L^{(1)}\wedge    L^{(3)} ) , \quad 
\ee
in which as before $L= g^{-1} dg$ (but with the group element depending on both bosonic coordinates and fermions) and the inner product is now the super-trace.  This is integrable with Lax \cite{Bena:2003wd} given by 
\be
 {\mathcal L}_\pm(z) = L^{(0)}_\pm + z  L^{(1)}_\pm + z^{\mp 2} L^{(2)}_\pm  + z^{-1}  L^{(3)}_\pm   \ , ~~~ z \in \mathbb{C} \ . 
\ee
The deformed theory \cite{Delduc:2013qra} reads 
\be
S_\eta  =  - \frac{1}{2 t}  \int d^2\sigma (\eta^{\mu\nu} - \epsilon^{\mu \nu}  )\langle L_\mu  , \Omega_\eta \cdot \frac{1 -\eta^2 }{\mathbbm{1} -\eta {\cal R}^g\cdot \Omega_\eta }  L_\nu   \rangle\ .   
\ee
The Lax has a similar structural form as in the undeformed case but is written in terms of currents 
\be
J_\pm = (\mathbbm{1} -\eta {\cal R}^g\cdot \Omega_\eta)^{-1} \cdot L_\pm \, , \quad \tilde{J}_\pm = (\mathbbm{1} + \eta {\cal R}^g\cdot \Omega^T_\eta)^{-1} \cdot L_\pm \ . 
\ee
The detailed combinations of how these enter into the Lax can be found in \cite{Delduc:2013qra}.  The $\eta$-deformed theory on this semi-symmetric space admits a $q$-deformed symmetry with real quantum group parameter 
\be
q  = \exp \left(t \varkappa  \right) \ , \quad \varkappa = \frac{2 \eta}{1-\eta^2}  \, . 
\ee
  Applied to the $AdS_5$ superstring with the $\frak{psu}(2,2|4)$ algebra   there is a natural proposal \cite{Hoare:2011wr} for the full quantum $\mathbb{S}$-matrix that respects the  q-deformed symmetries; it is given by a tensor product of two copies of the q-deformed $\frak{psu}(2|2)$ S-matrix blocks introduced in \cite{Beisert:2008tw}.
By matching this proposed  $\mathbb{S}$-matrix to the tree-level perturbative bosonic  S-matrix elements in the large tension limit, very compelling evidence was given in  \cite{Arutyunov:2013ega} that the $\eta$-deformation provides a Lagrangian description for this scattering theory.

 \subsection{A comment on modified supergravity }  

Starting with the type IIB superstring on $AdS_5 \times S^5$  one can evaluate the $\eta$-deformation as described above.  A serious puzzle at the time was that for a certain choice of ${\cal R}$-matrix the geometry and RR fields extracted from the action do not solve the conventional equations of motion of type II supergravity \cite{Arutyunov:2015qva}.\footnote{Since Lie super algebras admit inequivalent Dynkin diagrams there can be inequivalent choices of ${\cal R}$-matrices.  In developments subsequent \cite{Hoare:2018ngg} to this conference  it was shown  that for the    $\frak{psu}(2,2|4)$  Dynkin diagram with all simple roots fermionic, the unimodularity condition is obeyed and that the background fields of the  $\eta$-deformation can  be extracted and do solve conventional supergravity.  } 

It later transpired that this geometry, and others like it, solve a particular modification of supergravity \cite{Arutyunov:2015mqj} that depends on a distinguished Killing vector $I$ (detailed in appendix \ref{sec:mSUGRA}).   Algebraically, a necessary and sufficient condition for the $\eta$-model to have a standard supergravity \cite{Borsato:2016ose} solution is that the   ${\cal R}$-matrix is {\em unimodular} i.e. 
\begin{equation}
 I^c \equiv  f_{ab}{}^c  \mathtt{r}^{ab} = 0 \, ,  
\end{equation}
which has the interpretation that the adjoint representation of $\frak{g}_{\cal R}$ is traceless.  Historically the first solution of this {\em modified supergravity}  (sometimes called {\em generalised supergravity}), although not coined as such at the time, was obtained in 1993 in \cite{Gasperini:1993nz} via performing non-Abelian T-duality in of non-unimodular isometry group for which $f_{ab}{}^b$ is not zero (the more modern understanding of these backgrounds in terms of modified supergravity was provided in \cite{Hong:2018tlp}).  In general one anticipates that the non-Abelian dualisation for non-unimodular groups is afflicted with anomalies \cite{Elitzur:1994ri,Alvarez:1994np}.  

How should this be interpreted? The most obvious feature of modified supergravity is that there is no natural dilaton field, and so no obvious Fradkin-Tseytlin term.  This has prompted the suggestion that the corresponding worldsheet theories are not locally Weyl invariant (though is $\kappa$-symmetric and expected to be scale invariant).

However this is perhaps too strong a conclusion.  For instance centrally extended non-Abelian T-duality, which would seem to suffer from this problem, should  give rise to bone-fide supergravity solutions and consistent string theories at least when  it is equivalent to TsT transformations.  In   \cite{Wulff:2018aku} a resolution to this tension was suggested by proposing a non-local field redefinition that could remove the anomalous terms.  An inventive proposal \cite{Sakamoto:2017wor} is that the Weyl anomaly should be canceled with a modified Fradkin-Tseytlin term involving a dilaton field that is linear coordinates  $\tilde{x}$ that are T-dual to those adapted to the Killing vector $I$. This idea is quite natural from the perspective of the doubled world sheet but at first sight this seems to be unsatisfactory since the counter term looks to be non-local in the original field variables.  However the special properties of curvature tensors in two-dimension have been exploited to show that this counter term can be recast, after a suitable integration by parts, as a local counter term \cite{Fernandez-Melgarejo:2018wpg}.  Recently this point of view has been reinforced; in  \cite{Muck:2019pwj}   a local counter term is proposed for the Green-Schwarz string even when the target space obeys modified supergravity equations.    

To close this section let us remark on one nice direction, the models built on solutions of the Yang-Baxter equation (modified or not) display an elegance when transformed under the open-closed map \cite{Seiberg:1999vs} from the metric $G$ and $B$-field to open string metric $G_0$ and non-commutativity parameter $\Theta$: 
 \be
 G_0 + \Theta  = ( G+ B)^{-1} \, . 
 \ee 
 As first observed in \cite{vanTongeren:2016eeb} (for CYBE) and later extended \cite{Araujo:2017jap,Araujo:2017enj,Araujo:2018rbc} (including for mCYBE),    once appropriately dressed with Killing vectors (e.g. those corresponding to the directions of a TsT in the $c=0$ case)  one has that the non-commutativity parameter  $\Theta \sim \mathtt{r}$,    and moreover the divergence of the  non-commutativity parameter is the Killing vector $I$ that defines the modified supergravity \cite{Araujo:2017enj}. 
 
%
%
%
%
%
%
%
%
%
%
%
%
%
%
%
%
%
%

\section{$\lambda$-models} 
 
 We now turn to something seemingly quite different - the $\lambda$-models introduced by Sfetsos \cite{Sfetsos:2013wia}.  The initial motivation for these theories was to regulate the non-compactness of geometries produced by non-Abelian T-duality by embedding them as a patch of some larger manifold.  The idea was to construct  a one-parameter family of non-linear sigma-models, labelled by $\lambda \in [0,1]$, that interpolate smoothly between a CFT with a well understood geometric target space and the non-Abelian T-dual of a sigma model.  The actual outcome is probably even more valuable:   $\lambda$-models provide a wide class of novel integrable models that encode quantum group symmetries at the root of unity.
 
\begin{figure}
      \[   \begin{tikzcd}[column sep=small]
\textrm{WZW}~~G &\longrightarrow  & \textrm{nAbT-PCM}~~G\\  
  \lambda =0    & \longrightarrow & \lambda \sim 1   \\ 
  \textrm{gWZW}~~G/H &  \longrightarrow & \textrm{nAbT-PCM}~~G/H 
  \end{tikzcd}\] 
\caption{Integrable interpolations for the case of group manifolds (top) and symmetric spaces (bottom).  Assuming that we are dealing with compact spacetimes the direction of the arrows matches the RG flow.  } 
\end{figure}

    In the simplest case we consider the PCM on a group manifold $G$ of radius $\kappa$, with the action defined in eq.~\eqref{eq:PCM}, and its non-Abelian T-dual $\widehat{S}_{nAbT}$.  The $\lambda$-model interpolates between the Wess-Zumino-Witten model on $G$ at level $k$ and the non-Abelian T-dual theory with the deformation parameter given as the ratio 
 \be\label{eq:lambdadef}
 \lambda = \frac{k}{\kappa^2 + k} \,   .
 \ee 
 The field content is given by a group element $g\in G$.  Taking $\lambda \to 1$ is expected to give the non-Abelian dual of the principal chiral model as can be (somewhat heuristically) argued by an expansion of the group element
 \be
 g = \mathbbm{1}+ \frac{i v^a  T_a}{k} + O(k^{-2} ) \, . 
 \ee
 The term in the $\lambda$-deformed theory that survives the limit $k\to \infty$ is precisely the non-Abelian T-dual action.  Quantum mechanically the coupling $\lambda$ runs from $\lambda = 0$ in the UV to $\lambda \to 1 $ in the IR.   
 These integrable interpolating theories extend to both $\mathbb{Z}_2$ graded symmetric spaces $G/H$ \cite{Sfetsos:2013wia,Hollowood:2014rla}, in which the corresponding UV CFT is a coset-CFT realised as a gauged WZW model,  and to $\mathbb{Z}_{4}$ semi-symmetric spaces with application to the $\frak{psu}(2,2|4)$ superstring \cite{Hollowood:2014qma}.      
 
 We now proceed to discuss these interesting integrable models paying greatest attention to the simplest case of the theory on bosonic group manifolds. 
 
  \subsection{$\lambda$-deformations of the PCM on a group}
  
  For both the $\lambda$-theory on a group and on a $\mathbb{Z}_2$ graded symmetric space, Sfetsos \cite{Sfetsos:2013wia}  proposed an elegant modification of the Buscher procedure to construct, instead of a dual model, a  family of integrable models.\footnote{The case of the $\mathbb{Z}_4$ graded super-coset the Sfetsos procedure, at least in the form understood here, is not applicable.}   We start with the PCM on $G$ with radius $\kappa$ as defined in eq.~\eqref{eq:PCM} for a group element $\tilde{g}$,  
   \begin{equation}\label{eq:PCMaction}
 S_{\text{PCM}}[\widetilde{g}] = -  \frac{\kappa^2}{\pi} \int d^2\sigma  \, \langle \widetilde{g}^{-1}\partial_+ \widetilde{g}  \widetilde{g}^{-1}\partial_- \widetilde{g} \rangle \, , 
 \end{equation}
 in which $\langle \cdot, \cdot \rangle$ is the Killing form.   As with the   Buscher procedure we begin by gauging the $G_L$ action
\be
 G_{L} :    \tilde{g}\rightarrow h^{-1}\tilde{g} \, ,
\ee   
by promoting partial derivatives in the PCM to covariant  derivatives,  $  D= \partial + A$, with   gauge fields $A =i  A^a T_a$ that  transform as 
  \begin{equation}
  A \rightarrow h^{-1} A h - h^{-1} \mathrm{d} h \ .
  \end{equation}
The gauged PCM is then
     \begin{equation}\label{eq:PCMaction}
 S_{\text{gPCM}}(\widetilde{g}, A) = -  \frac{\kappa^2}{\pi} \int  d^2\sigma \, \langle \widetilde{g}^{-1}D_+ \widetilde{g} ,  \widetilde{g}^{-1}D_- \widetilde{g} \rangle \, . 
 \end{equation}

Instead of adding a Lagrange multiplier term as in the derivation of the non-Abelian T-dual, let us now add in a new sector given by a Wess-Zumino-Witten (WZW) model on $G$ at level $k$ for a different group element $g\in G$.  We recall that the WZW action is given by   \cite{Witten:1983ar}   
 \be \label{eq:WZWaction1}
  S_{\text{WZW,k}}(g) = -\frac{k}{2\pi}\int_\Sigma  d^2\sigma \,  \langle g^{-1} \partial_+ g , g^{-1} \partial_- g \rangle - \frac{  k}{4\pi }       \int_{M_3}  H   \ , 
   \ee
   where $H$ is the closed 3-form (locally satisfying $H = \mathrm{d}B$) given by
   \begin{equation}
 H = \frac{1}{6}   \langle   \bar g^{-1} d\bar g, [\bar g^{-1} d\bar g,\bar g^{-1} d\bar g]  \rangle ,
   \end{equation}
 with $\bar{g}$ the extension of $g$ into $M_3 \subset G$ such that $\partial M_3 = g(\Sigma)$ (we assume that  $H_2(G)=0$ and for simplicity we take $G$ to be semi-simple).  To ensure that the path integral doesn't depend on the choice   of extension,  the third cohomology class $[H]/2\pi\in H^3(G)$ must be integral and so, if this class is non-empty,  the level $k$ should be quantised ($k\in \mathbb{Z}$ for $SU(N)$).  This theory is conformal and   characterised by a $G(z) \times G(\bar{z})$  current algebra\footnote{We switch to Euclidean signature with  $z = x^0 + i x^1 = i \sigma^+$ and $\bar{z}=x^0 - i x^1 = i \sigma^-$  and   $(x^0,x^1) = (i \tau, \sigma)$.}.   From the OPE of the energy-momentum tensor, given via the Sugawara construction, one finds the central charge 
 \begin{equation}
c = \frac{k \,\text{dim}(G)}{k+h^\vee }\ ,
\end{equation}
 where $h^\vee$ is the dual Coxeter number of $G$.   
  
On this WZW we will also gauge a subgroup of the global symmetry.  The gauging of symmetries in the WZW model is a little delicate (see \cite{Witten:1991mm} and \cite{Hull:1989jk})  but one can gauge the diagonal symmetry\footnote{A generalisation is consider asymmetric gauging constructed by means of an algebra automorphism. When this automorphism is outer  this can result in a different theory.  This has recently been employed to construct the context of $\lambda$-models \cite{Driezen:2019ykp}.    } acting as 
\be
 G_{\text{diag}} :    g\rightarrow h^{-1}g h\, .
\ee  
 by  replacing the WZW model \eqref{eq:WZWaction1} with the $G/G$ gauged WZW model \cite{Bardakci:1987ee,Gawedzki:1988hq} 
  \begin{equation}
  S_{\text{gWZW,k}}(g,A) = S_{\text{WZW},k}(g) + \frac{k}{\pi} \int  d^2\sigma   \, \langle A_- , \partial_+ g g^{-1} \rangle - \langle A_+ ,  g^{-1}\partial_- g \rangle  + \langle A_-, g A_+ g^{-1} \rangle - \langle A_- , A_+ \rangle \,    .
  \end{equation}
Note here that  we have used the {\em same  gauge field} to gauge the $G_L$ of the PCM and the $ G_{\text{diag}}$ of the WZW.   We consider then the sum
\be
S_{k,\lambda}(g,\tilde{g}, A)  = S_{\text{gWZW,k}}(g,A) + S_{\text{gPCM}}(\tilde{g}, A)
\ee
 in which the two terms are effectively coupled through the  gauging. 
 Finally, we can fix the gauge to $\widetilde{g} = 1$ yielding 
\begin{equation}
  \begin{aligned}\label{eq:LambdaAction1}
  S_{k,\lambda}(g, A) = &\, S_{\text{WZW},k} (g) 
  - \frac{k}{\lambda \pi} \int d^2\sigma \langle A_+  , O_{g^{-1}}^{-1} A_- \rangle
  + \frac{k}{\pi} \int d^2\sigma \, \langle A_- , \partial_+ g g^{-1} \rangle  -  \langle A_+ , g^{-1}\partial_- g   \rangle   \,,
  \end{aligned}
  \end{equation}
where we introduced the useful operator  
\begin{equation}
 O_g = \left(\mathbbm{1} - \lambda D \right)^{-1} \, , 
\end{equation} 
given in terms of the adjoint action $D(T_a) =\mbox{ad}_gT_a= g T_a g^{-1}$ and  $\lambda$ is the  parameter given by eq.~\eqref{eq:lambdadef}.  As with the Buscher procedure, the gauge fields are non-dynamical and can be integrated out and replaced with their on-shell values\footnote{At the quantum level one anticipates that this procedure would generate additional contributions perturbatively in $\frac{1}{k}$, and understanding the implication of this for integrability is an open question. } 
\begin{equation}\label{eq:GaugeConstraints}
A_+ = \lambda \,O_g  \partial_+ g g^{-1}\, , \quad   
A_-  = - \lambda \,O_{g^{-1}} g^{-1}  \partial_- g\,  ,
\end{equation} 
  resulting in the action 
\begin{equation}
 \begin{aligned}\label{eq:LambdaAction2}
S_{k,\lambda}(g) &= S_{\text{WZW},k}(g) + \frac{k \lambda}{\pi}\int d^2\sigma \, R_+^a \left(O_{g^{-1}} \right)_{ab} L_-^b \,  , \nonumber  
\end{aligned}
\end{equation}
where $R_\pm = \partial_\pm g g^{-1}$ and $L_\pm = g^{-1} \partial_\pm g$.  
%
%
It is straightforward to read off from this action the target space data which can be expressed  in terms of the left-invariant forms $L^a$  as:
\begin{eqnarray}
\begin{aligned}
\mathrm{d}s^{2}_\lambda &= k \left( O_{g^{-1}} + O_g -  \eta \right)_{ab} L^a\otimes L^b  \ ,  \label{eq:MetricLamba1}\\
B_\lambda  &= B_{\text{WZW}} + \frac{k}{2} \left( O_{g^{-1}} - O_g \right)_{ab} L^a \wedge  L^b \label{eq:BfieldLambda} \ , 
\end{aligned}
\end{eqnarray} 
where we have used that $R^a = D^a{}_{b} (g) L^b$   and in which $B_{\text{WZW}} $ is a local potential such that  $\mathrm{d}B_{\text{WZW}} = \frac{k}{6} F_{abc} L^a \wedge L^b \wedge L^c $. 
 In addition the Gaussian elimination of the gauge fields, when performed in a path integral, results in a non-constant dilaton profile,
\begin{equation}\label{eq:DilatonProduced}
\Phi = \Phi_0 -\frac{1}{2} \ln \det O_{g^{-1}} \ ,
\end{equation}
in which constants are absorbed into $\Phi_0$.

{\bf Example:} Let us consider the case of  $G= SU(2)$ with the group element  parametrised by 
\be
g= \begin{pmatrix}
\cos \alpha + i \sin \alpha \cos \beta  & \sin \alpha \sin \beta e^{-i\gamma}  \\   -\sin \alpha \sin \beta e^{i\gamma} & \cos \alpha - i \sin \alpha \cos \beta 
\end{pmatrix} \, . 
\ee
The corresponding $\lambda$-deformed target space data  is given \cite{Sfetsos:2013wia} by
\be
\begin{split}
\hskip -2 cm
\mathrm{d}s^{2}_\lambda & = k \left(\frac{1+\lambda }{ 1-\lambda} d\alpha^2 + \frac{1-\lambda^2}{ \Delta(\alpha)} \sin^2\alpha \ ds^2(S^2) \right)\ ,
\\
& B_\lambda  = k\left( - \alpha + \frac{(1-\lambda)^2}{\Delta(\alpha)} \cos\alpha \sin \alpha \right)  {\rm{Vol}}(S^2)\, ,\\ 
& e^{-2 \Phi}  = \Delta(\alpha) \, ,
\end{split}
\label{su2def}
\ee
where 
\be
\Delta(\alpha) = (1-\lambda)^2 \cos^2\alpha + (1+\lambda)^2 \sin^2 \alpha\, ,
\ee
and $ds^2(S^2) = d\beta^2 + \sin^2\!\beta\ d\gamma^2$ and ${\rm{Vol}}(S^2) =\sin\beta \ d\beta \wedge d\gamma$.  The study of how this, and more general $\lambda$-model backgrounds can be embedded to supergravity solutions with appropriate RR fields was commenced in \cite{Sfetsos:2014cea,Demulder:2015lva} with a first principles derivation starting from the $\lambda$-deformation of the GS superstring on semi-symmetric spaces    given in \cite{Borsato:2016ose}.
  \subsubsection{$\lambda$-model as an RG flow}
  
It is helpful to consider the limiting cases of $\lambda \sim 0$ and $\lambda \to 1$.  First, for small deformations 
$\lambda = \epsilon \ll 1$ we have that 
\begin{equation}
S_{k,\lambda}(g) \sim  S_{\text{WZW},k}(g) +  \epsilon \int  \sum_{a} R_+^a L_-^a  \, ,
\end{equation}
i.e. a current-current perturbation of a CFT.  Notice importantly that we sum over all directions in the algebra in this perturbation - this is not a marginal deformation, instead it is (for compact groups) a marginally relevant operator that triggers an RG flow.  So we are considering here an integrable QFT whose UV completion is the WZW CFT.   This is borne out by the beta function of $\lambda$. Calculated in \cite{Itsios:2014lca} for the full theory, eq.~\eqref{eq:LambdaAction2}, to all orders in $\lambda$ (but leading in $k^{-1}$)  one has
\begin{equation}
\mu \frac{d}{d \mu} \lambda = - \frac{c_G}{k} \frac{\lambda^2}{(1+ \lambda)^2} \, , 
\end{equation}
with the solution 
\be
\log \lambda^2 + \lambda - \frac{1}{\lambda}  = -  \frac{c_G}{k} \log\mu/\mu_0 \, . 
\ee
As $\lambda \to 1$ we are moving towards to IR of the theory.  Some intuition can be gained by expanding the action around this point, however care must be taken.  In a correlated scaling limit in which we take $k \to \infty$ and expand the group element
\be\label{eq:scaling}
g   = \mathbbm{1}+ \frac{i}{k} v^a T_a + O(k^{-2}) \, ,
\ee
the leading contribution to the $\lambda$-geometry in eqs~\eqref{eq:MetricLamba1}-\eqref{eq:DilatonProduced} 
reduces exactly to that of the $G_L$ non-Abelian T-dual of the PCM.   The reason for this is that in the limit, the entire gauged WZW model  reduces to the Lagrange multiplier term enforcing a flat connection, i.e. the Sfetsos procedure reduces to a Buscher procedure \cite{Sfetsos:2013wia}.   Though this connection is quite enticing, it is not clear that this scaling limit described by eq.~\eqref{eq:scaling} provides a robust quantum argument that the IR of the theory is described by the non-Abelian dual. We will return to this shortly and provide further evidence.  However assuming it to be true the picture one might have in mind is that an RG flow is triggered from the UV, but in the IR admits a description that is dual to the PCM described by the exact S-matrix eq.~\eqref{eq:PCMSmatrix}.  

When considering $\lambda$-deformations of a non-compact group one finds that the beta function occurs with an opposite sign.  This suggests that   by performing a $\lambda$-deformation simultaneously to a compact space and a non-compact space one could obtain a marginal deformation.  Indeed, this idea paved the way to the embedding of $\lambda$-deformations as supergravity solutions at first by adjoining an appropriate RR sector to the above bosonic construction  \cite{Sfetsos:2014cea,Demulder:2015lva} and  subsequently in the full Green-Schwarz string \cite{Borsato:2016ose}.  This is supplemented  the understanding that $\lambda$ is indeed one-loop marginal \cite{Appadu:2015nfa} in the case relevant to $AdS_5 \times S^5$.

 \subsubsection{Lax Formulation}

Classical integrability   is  elegantly demonstrated by specifying a Lax  connection given in terms of the gauge fields subject to their on-shell identification \cite{Hollowood:2014rla}  recovering the result first given in a direct fashion by   \cite{Sfetsos:2013wia}.  Indeed, the   equation of motion for the group element $g$  can be rewritten as 
\begin{equation}\label{eq:LambdaEOM}
\partial_\pm A_\mp = \pm \frac{1}{1+\lambda}\,\left[ A_+ , A_ - \right]\, .
\end{equation}
Hence, effectively we have recast the second order equation for the field $g$  as two first order equations for $A_\pm$ subject to the constraints of eq.~\eqref{eq:GaugeConstraints}. The equations for $A$ follow from demanding that the  Lax connection 
\begin{equation}\label{eq:LaxLambda}
{\mathcal L}_\pm(z) = -\frac{2}{1+\lambda}\, \frac{A_\pm}{1 \mp z}\, , \qquad z \in \mathbb{C}\, ,
\end{equation}
 satisfies the flatness condition $\mathrm{d} {\mathcal L} + {\mathcal L} \wedge {\mathcal L} = 0$ together with the identification  of eq.~\eqref{eq:GaugeConstraints}.

  \subsubsection{Multi-parameter $ \lambda$-models}
 In passing we remark that by following the same Sfetsos procedure starting with the most general PCM depending on a matrix $E_{ab}$ as in eq.~\eqref{eq:genPCM}, what results is a theory that depends on an  matrix of deformation parameters 
\be \lambda^{-1}_{ab} =  \frac{1}{k} \left( E + k \mathbbm{1}\right)_{ab} \, .  \ee
One might expect this to yield an integrable $\lambda$-model  when the starting PCM also happens to be integrable.  For example, performing the Sfetsos procedure with the $\eta$-deformed PCM, i.e. choosing $E= \left(  \mathbbm{1} -    \eta {\cal R}\right)^{-1}$,  does indeed give rise to a two-parameter anisotropic $\lambda$ deformation of the WZW model for which an explicit Lax can be obtained. This kind of anisotropic, or XXZ,  $\lambda$-model has been studied in  \cite{Sfetsos:2014lla,Sfetsos:2015nya,Chervonyi:2016bfl,Appadu:2018jyb}.   Other more general types of $\lambda$-models including products of interacting WZW factors have been developed in a sequence of works \cite{Georgiou:2016urf,Sfetsos:2017sep,Georgiou:2018gpe,Sagkrioti:2018rwg}.

 \subsubsection{ $\star$ Boundaries and D-branes  $\star$}

Identifying precisely the D-brane states of a CFT realised as  a sigma-model in a curved background is challenging.  For the WZW model on a compact group however this can be answered precisely.  Classically one finds boundary conditions in which left and right currents are glued with the geometric interpretation of  D-branes  wrapping conjugacy classes. Flux quantisation   restricts these conjugacy classes to be those labeled by integral highest weight representations \cite{Alekseev:1998mc,Gawedzki:1999bq,Felder:1999ka,Stanciu:1999id}. For example for the $SU(2)_k$ WZW there are D0 branes at $g=e$ and $g= - e$ and then  $k-1$ (Euclidean) D2-branes wrapping   $S^2$ conjugacy classes.    These conjugacy classes can be further twisted by the action of an algebra automorphism, $\Omega$, leading to D-branes wrapping {\em twisted}-conjugacy classes  defined as 
\be
C_{\Omega}(g) = \{ h g \omega( {h^{-1}}) | \forall h \in G \} \, , \quad \omega(\exp x) =\exp \Omega(x) \, .  
\ee

On general grounds D-branes (and boundaries) break symmetries.  An obvious question to ask is: viewing the $\lambda$-model as a deformation of the WZW can we identify D-brane configurations for which integrability persists?  The technique used in \cite{Driezen:2018glg} was to find conditions such that a monodromy object $T_b(z)$ formed by taking the Wilson line of the Lax to the boundary and reflecting back,  obeys $\partial_\tau{T}_b(z) = [T_b(z), N]$ for some matrix $N$ and such that $\Tr{T}_b(z)$  generates conserved charges.  For the $\lambda$-model this is the case when 
\begin{equation}
O_{g^{-1}} L_- |_{\partial \Sigma}  = - \Omega\cdot O_{g}  R_+ |_{\partial \Sigma}\, ,
\end{equation}
with $\Omega$ a constant involutive and metric preserving algebra automorphism.  The $\lambda$ dependence in this formula is just right to compensate for that appearing in the metric; the D-branes still define twisted conjugacy classes with the additional restriction that the automorphism defining the twist is  involutive. 

Returning to the example of $G= SU(2)$.  Here there are still D2 branes wrapped on the $S^2$ located at values of $\alpha = \frac{n \pi }{k}$, $n=0,  \dots k$ and the effect of the deformation is to, in essence, determine the size of these cycles. Although  the flux $H$ is sensitive to the deformation, so to is the  field strength living on the $D2$ world volume. These are compatible such that the flux quantisation condition is independent of the deformation parameter (as it should be!).   One can   analyse the fluctuations of the DBI action world volume. There is an s-wave fluctuation of position that is massive for all values of $\lambda$ -- this is the flux stabilisation in another guise.  For $\lambda =0$ there is a  p-wave triplet of position fluctuations  corresponding to Goldstone modes associated to the brane breaking an $SU(2)$ global symmetry of the target space (selecting an $S^2 \subset S^3$).  For  $\lambda \neq 0$  there is no longer such an $SU(2)$  symmetry to be broken by the brane and these modes acquire a mass.  
 
 \subsubsection{ $\star$ $\lambda$-models, Quantum Inverse Scattering and ${\mathbb S}$-matrices $\star$} 

Thus far we have primarily considered classical integrability with only hints as to the quantum integrable structure.  An important question is if integrability persists at a quantum level and can the integrable structure be quantised?  This is   a very delicate matter; the classical dynamics of the $\lambda$-model can be recast in terms of two commuting classical Kac-Moody current algebras
\be
\{ {\cal J}^a_\pm(x) , {\cal J}^b_\pm(y) \}_{p.b.} = f^{ab}{}_c {\cal J}_\pm(y) \delta(x-y) \pm \frac{k}{2\pi} \delta^{ab} \delta^\prime (x- y) \, ,  
\ee
in which the currents can be related to the on-shell value of gauge fields via
\be
{\cal J}_\pm = - \frac{k}{2\pi} \left( \frac{1}{\lambda} A_\pm - A_\mp \right) \, .
\ee
It is clear then that the Lax connection  in turn  can be cast in terms of these currents.  One should like to lift this bracket to the monodromy operator, $T(z)$,  generating the conserved charges, and then by regularizing the theory on a spatial lattice perform a quantisation.  However, the non-ultra-locality  (the Schwinger term / derivative of the delta function)   prevents this from being done in any straightforward way.  

An alternative philosophy, adopted in the $\lambda$-models in \cite{Appadu:2017fff,Appadu:2018jyb},  is to start directly with a bare theory on a lattice defined by an integrable Hamiltonian.  One can exploit e.g. Algebraic Bethe Ansatz techniques \cite{Faddeev:1996iy} to determine the ground state and excitations.   With a carefully chosen continuum limit one can then obtain relativistic dispersion relations and an S-matrix compatible with the axioms of factorised scattering of an integrable QFT.  Armed with such an S-matrix one can calculate e.g. the free energy at finite chemical potential or high temperature, and compare to the perturbative result coming from the QFT.  In a sense this approach is to take a {\em classicalisation} of a quantum theory rather than a {\em quantisation} of a classical theory.  

One may still wish to motivate the bare lattice theory directly from the classical field theory. An approach to this is Faddeev-Reshetikhin \cite{Faddeev:1985qu} alleviation idea, which in this case \cite{Appadu:2017fff} consists of taking a delicate limit of $k\to 0$ and $\lambda \to 0$ whilst keeping fixed
\be
\nu = \frac{k}{4\pi \lambda} \, . 
\ee 
In this limit the theory is no-longer the one we started with,  the central term drops and the quantisation proceeds by promoting the Poisson algebra to an operator algebra  on a light cone lattice of spacing $\Delta$. On a fixed time slice the currents living on the null links of this lattice, labelled by their direction $\pm$ and location $x=n\Delta $,    obey an operator algebra
 \be
[{\cal J}_{\pm,m}^a , {\cal J}_{\pm,n}^b]  = \frac{i}{\Delta} f^{ab}{}_c  {\cal J}_{\pm,n}^c \delta_{nm}  \, . 
\ee 
One can then choose, with post-hoc justification, to realise this algebra by generators in a symmetric rank $k$ representation. 

Crucial to an integrable lattice model is the (quantum) ${\mathbb R}$-matrix which obeys a (quantum)  Yang-Baxter equation with spectral parameter, which in the semi-classical limit should correspond to the transport of the Lax connection along a null link \cite{Destri:1987ug}.    This information is sufficient to uniquely determine the quantum integrable lattice model. After applying Algebraic Bethe Ansatz, the eigenstates of lattice theory are determined by a spin chain.  For $G= SU(2)$ (which we assume for the rest of this section) this spin chain is the famous Heisenberg anti-ferromagnetic $ XXX$ spin-chain but with a) spin $k/2$ and b) alternating inhomogeneity determined by the parameter $\nu$ introduced above.  

In the thermodynamic limit the ground state is determined by a density, in the space of rapidities $z$, of solutions of the Bethe equations called $k$-strings.\footnote{These are a condensate of $k$ solutions to the Bethe equations centred around $z$ with imaginary parts offset by increments of $k /2$.}  More pertinent are the excitations above the ground state. These spinon excitations carry a hidden kink structure in that they are associated to a set of $k+1$ vacua corresponding to the integrable highest weights of $SU(2)$ at level $k$.\footnote{The situation for $SU(N)$ is more involved with kinks carrying an addition label that captures the rank of the group.} One can then calculate an effective kernel, ${\cal K}$, that describes the scattering of these kinks off each other as well as a dispersion relation between the spinon energy, $\varepsilon(z)$, and momenta,  $p(z)$, given by
\be
\sinh^2 \pi \nu \tan^2 \frac{ \varepsilon(z)\Delta }{2} - \cosh^2 \pi \nu \tan^2 \frac{p(z) \Delta   }{2}  = 1 \, .
\ee
Due to the inhomogeneity $\nu$ these excitations are gapped.   This is not, of course, a relativistic dispersion relation however there is a continuum limit $\Delta \to 0 $ {\em and} $ \nu \to \infty$   which, after identifying  rapidity as $\theta = z \pi$, yields a relativistic dispersion 
\be
p = m \sinh \theta \ , \quad \varepsilon = m \cosh \theta \ ,  \quad \varepsilon^2 - p^2 = m^2 \ , \quad~~\textrm{with} ~~ \frac{m}{4} = \frac{1}{\Delta} \exp( - \pi \nu) ~ \textrm{fixed.} 
\ee
The mass here is generated via transmutation from the cut-off $\mu = \Delta^{-1}$ and the dimensionless parameter $\nu$.  A first consistency check comes from demanding how $\nu$ must be changed with scale $\mu$ such that $m$ is fixed; this returns exactly the beta function for $\lambda$ in the UV limit.   The spin chain TBA equations that determine the above ground state involve a particular scattering kernel.  The resolvent of this in turns determines the ${\mathbb S}$-matrix element $S(\theta)$ describing  scattering of the spinon excitations above the ground state
\be
\frac{1}{2 \pi i } \frac{d S(\theta) }{ d  \theta} =  \delta(\theta)  - \int_0^\infty \frac{dx}{\pi} \cos(x \theta) \hat{R} (x) \ , \quad R(x) =  \left( 1+ \coth(k \pi x/2) \right) \tanh (\pi x /2)   \, . 
\ee
This scattering kernel can be realised as coming from an integrable relativistic ${\mathbb S}$-matrix that consists of two factors, a universal $SU(2)$ (Yangian) invariant S-matrix block $S^{SU(2)}(\theta)$  that we saw already in the case of the PCM,  and a factor that encodes the kink structure  $S^{RSOS}_k(\theta)$:
\be
{\mathbb S} = X(\theta)  S^{SU(2)}(\theta) \otimes S^{RSOS}_k(\theta) \, , 
\ee 
where as with the PCM the factor $X(\theta)$ is required to ensure the completed ${\mathbb S}$-matrix has the correct analytic structure.  The kink part of the ${\mathbb S}$-matrix has a quantum group at root of unity structure with\footnote{Note the subtle shift in the level, a familiar feature from quantum treatments whose appearance here is determined by BA equations in the thermodynamic limit.  }  
\be
q  = \exp( - i \pi  / ( k+2) )  \ . 
\ee
As $k \to \infty$ this  $S^{RSOS}_k(\theta)$ factor becomes, after a change of basis (vertex-face map) an $SU(2)$ Yangian  invariant block and the entire ${\mathbb S}$-matrix matches that of the PCM.  This is expected to be the ${\mathbb S}$-matrix equivalent statement of $k \to \infty$ scaling limit of the $\lambda$-model in which the non-Abelian T-dual is reproduced. 

In  \cite{Appadu:2017fff} this analysis was similarly carried out for the $SU(N)$ equivalent $\lambda$-deformation, providing a derivation of a previously conjectured ${\mathbb S}$-matrix.  The  free energy calculated from this at high temperatures (driving the system to the UV) matches that of the central charge of the WZW model,  and at $T=0$ but at finite chemical potential a precise matching can be made with perturbation theory including a full recovery of the $\beta$-function of the coupling $\lambda$.  In \cite{Appadu:2018jyb} a similar analyse was conducted for two-parameter anisotropic $\lambda$-deformations in the case of $SU(N)$ making a link to an $XXZ$ spin $k/2$ spin chain. 

For the more general cases of symmetric spaces and semi-symmetric spaces the procedure described above has not yet been implemented but forms of the  ${\mathbb S}$-matrices are conjectured   \cite{Hollowood:2015dpa}.

 \subsection{ $\star$  $\lambda$-deformations of symmetric and semi-symmetric spaces $\star$} 

We can also consider $\lambda$-type deformations applied to a symmetric space $G/H$.  Here we begin again with the Sfetsos procedure starting with the  symmetric space sigma model (SSSM) for a group element $\tilde{g}\in G$ that is invariant under a local $H_R$ action $\tilde{g} \to \tilde{g} h$.  It is convenient here to construct this theory with some gauge fields  $B_\pm \in \frak{h}$ that transform as $B_\pm \to h^{-1} ( B_\pm + \partial_\pm) h$ and adopt the action 
\be
S_{\textrm{SSSM}, \kappa^2}[\tilde{g} , B] = -\frac{ \kappa^2}{\pi } \int d^2 \sigma \langle \tilde{g}^{-1} \partial_+ \tilde{g} - B_+,    \tilde{g}^{-1} \partial_- \tilde{g} - B_- \rangle \, . 
\ee 
We augment this by introducing $\dim G$ extra degrees of freedom with a WZW model $S_{\textrm{WZW},k} [g]$ for a group element in $g\in G$.  The WZW and the SSSM are coupled by gauging a common $G$ action so that the total local symmetries are 
\be
\tilde{g} \to \hat{g}^{-1} \tilde{g}   h \ , \quad   g \to \hat{g}^{-1} g   \hat{g} \ , \quad \hat{g} \in G \ , ~~ h \in H \ . 
\ee
As before this gauging is done by introducing a common gauge field $A_\pm \in \frak{g}$, using minimal coupling in the PCM and replacing the WZW with the $G/G$  gauged  WZW model.    We gauge fix $\tilde{g} = \mathbbm{1}$ and integrate out the gauge fields $B_\pm$ to yield   
\begin{equation}
  \begin{aligned}\label{eq:cosetLambdaAction1}
  S_{k,\lambda}(g, A) = &\, S_{\text{WZW},k} (g) 
  + \frac{k}{  \pi} \int d^2\sigma \langle A_- , \partial_+ g g^{-1} \rangle- \langle A_+ ,   g^{-1}\partial_-g  \rangle +  \langle A_- , g A_+g^{-1}    \rangle  -   \langle A_+ ,  \Omega(A_-)  \rangle 
    \end{aligned}
  \end{equation}
  in which the operator\footnote{Not to be confused with the gluing automorphism entering into the construction of D-branes.} $\Omega = \Omega^T= \ P^{(0)}+   \lambda^{-1} P^{(1)}$ involves the projectors on to the subgroup, $ P^{(0)}$, and coset,  $ P^{(1)}$.  The  action has a residual $H$ gauge symmetry under which $A_{\pm}^{(0)}$ transforms as connections and $A_{\pm}^{(1)}$ transforms adjointly.  Both these gauge fields can be integrated out to give constraints 
  \be\label{eq:Aconstraints}
  A_+ = - (D_g -\Omega^T)^{-1} \partial_+ g g^{-1} \ , \quad  A_- =  (D_{g^{-1} }-\Omega)^{-1} g^{-1} \partial_- g \, . 
  \ee
  Substituting these on-shell values into the action one finds the   final $\lambda$-deformation of the $G/H$ WZW \cite{Sfetsos:2013wia}: 
  \begin{equation}
  \begin{aligned}\label{eq:cosetLambdaAction2}
  S_{k,\lambda}(g ) = &\, S_{\text{WZW},k} (g) 
  + \frac{k}{  \pi} \int d^2\sigma \langle  \partial_+ g g^{-1}      , (\mathbbm{1} - D_g \Omega)^{-1}g^{-1}\partial_-g  \rangle \, . 
      \end{aligned}
  \end{equation}
  To interpret this as a sigma model and extract the target space geometry one should ensure that the choice of $g$ fixes the $H$ action, just as one would with a conventional $G/H$ gauged WZW.  There is also a dilaton profile produced in integrating out the gauge fields 
  \be
  e^{-2 \Phi} = e^{-2 \Phi_0} \det( D_g - \Omega) \, . 
  \ee
  The equations of motion can be written as 
  \be
  [\partial_+ - A_+ , \partial_- - \Omega(A_-)] = 0 \, , \quad   [\partial_+ - \Omega(A_+) , \partial_- - A_-] = 0 \, . 
  \ee
  The Lax connection is given by 
  \be
   {\cal L}_\pm(z) = - A^{(0)}_\pm - z^{\pm 1} \lambda^{-1/2}  A^{(1)}_\pm \, . 
   \ee
   with gauge fields subject to the identifications of eq.~\eqref{eq:Aconstraints}

  For  the case relevant to the $AdS_5\times S^5$ superstring we have  a $\mathbb{Z}_4$ graded super-algebra $\frak{g}$ with bosonic sub-algebra $\frak{h}= \frak{g}^{(0)}$.  Whilst a Sfetsos procedure does not quite hold here, the final result \cite{Hollowood:2014rla}  requires only small modifications to the action of eq.~\eqref{eq:cosetLambdaAction1}.      The group element,  $g$, is now understood to be an element of the supergroup and the inner product is the super-trace.  The  gauge fields $A_\pm$ are valued in the whole of $\frak{g}$ though only $A^{(0)}_\pm$  correspond to residual gauge symmetry.  The operator $\Omega$ is modified and given by 
  \be
  \Omega =  P^{(0)} + \lambda^{-1} P^{(1)} + \lambda^{-2} P^{(2)} + \lambda P^{(3)} \ , \quad   \Omega^T =  P^{(0)} + \lambda  P^{(1)} + \lambda^{-2} P^{(2)} + \lambda^{-1} P^{(3)} \, .
  \ee
 and the Lax is given by 
 \be
 {\cal L}_\pm(z) = -A_\pm^{(0)} - z \lambda^{\pm1/2} A_\pm^{(1)}- z^{\mp2} \lambda^{-1} A_\pm^{(2)} - z^{-1}  \lambda^{\mp 1/2} A_\pm^{(3)} \, . 
 \ee

  \subsection{$\star$ The $\lambda$-model in the Doubled Approach $\star$ }

The $\lambda$-deformed WZW model can also be derived from    a doubled worldsheet action given by eq.~\eqref{eq:Sdoubled} with some small modification \cite{Klimcik:1996nq,Klimcik:1996hp,Klimcik:2015gba} that relaxes the Drinfeld double structure and which have been dubbed ${\cal E}$-models.  These are defined on an algebra $\frak{d}$  (group $\mathbb{D}$) of dimension $2d$   equipped with an ad-invariant split signature inner product $\eta = \langle\langle \cdot , \cdot \rangle \rangle$.    We require that $\frak{d}$ admits a decomposition $\frak{d} = \tilde{\frak{h}} \oplus \frak{k}$ with $\tilde{\frak{h}}$ a maximally isotropic sub-algebra (with corresponding subgroup $\tilde{H}$) -- this is sometimes called a Manin quasi-triple.  Unlike in a Drinfeld double,  no requirement is placed on the complement $\frak{k}$ beyond the fact it be isotropic  - it need not be a sub-algebra and in general the coset $\mathbb{D}/\tilde{H}$ is neither a group nor a symmetric space.   On $\frak{d}$ we introduce an   self-adjoint operator ${\cal E}$ that squares to the identity and  that can be parameterised in terms of the generalised metric as ${\cal E}_{A}{}^B= {\cal H}_{AC} \eta^{C B}$.    In this more general setting we can still consider the doubled worldsheet action to be given by eq.~\eqref{eq:Sdoubled}.   From this one can eliminate halve the degrees of freedom by parametrising the group element on   $\mathbb{D}$ as $\mathbbm{g}(x, \tilde{x}) = m (x) \tilde{h}(\tilde{x})$ with $m(x)$ a representative of the coset $\mathbb{D}/\tilde{H}$.  The degrees of freedom contained in $\tilde{h}(\tilde{x})$ can be eliminated and what results is a conventional non-linear sigma-model on the target space  $\mathbb{D}/\tilde{H}$ given by 
 \be \begin{aligned}\label{eq:actEmod}
  S_{\mathbb{D}/\tilde{H}} &=  S_{WZW}[m] - \frac{1}{\pi} \int d\sigma d\tau  \langle\langle  P_{{\cal E}} (m^{-1} \partial_+ m) ,  m^{-1} \partial_- m  \rangle \rangle  \, , \\ 
  S_{WZW}[m] &= \frac{1}{2 \pi}  \int d\sigma d\tau   \langle\langle m^{-1} \partial_+ m,  m^{-1} \partial_-  m \rangle \rangle   + \frac{1}{24 \pi} \int_{M_3}   \langle\langle  m^{-1} d m , [ m^{-1} d m ,m^{-1} d m ]   \rangle \rangle   \, . \end{aligned}
\ee
Here the operator  $P_{{\cal E}}$ is constructed such that $\textrm{Im} P = \tilde{\frak{h}}$ and   $\textrm{Ker} P = (\mathbbm{1}+ \textrm{ad}_m \cdot {\cal E} \cdot  \textrm{ad}_{m^{-1}})\frak{d} $ \cite{Klimcik:2015gba}. Note that the action is evaluated on the coset representative $m(x)$ using the inner product on $\frak{d}$.   When   $\frak{k}$ also forms an isotropic subalgebra, we return to the case of a Drinfeld double;  the WZW part of the action drops out and one recovers exactly the form of Poisson-Lie $\sigma$-model encountered previously.   When   $\frak{k}$ is not a subalgebra,  unlike the case of Drinfeld double, it is not possible to reach a PL dual model by reversing the parametrisation, eliminating $m(x)$ and retaining $\tilde{h}(\tilde{x})$.
 
 The $\lambda$-deformed WZW on a group $G$  (algebra $\frak{g}$) is of this form \cite{Klimcik:2015gba}; one takes as the ``double'' $\frak{d} = \frak{g} \oplus \frak{g}$ with the inner product such that the diagonally embedded subgroup is a maximal isotropic.     In particular letting elements of $\frak{d}$ be represented by pairs $\{x,y\}$ with $x,y \in \frak{g}$ we have 
 \be
    \langle\langle \{x_1, y_1\} , \{x_2, y_2\} \rangle \rangle = \langle x_1, x_2 \rangle -  \langle y_1, y_2 \rangle \, , 
 \ee
 where $\kappa = \langle \cdot , \cdot \rangle$ is the Killing metric on $\frak{g}$.  The complementary space $\frak{k}$ of  elements of the form $\{x, -x\}$ is not a closed subgroup.   To obtain the $\lambda$-model action \eqref{eq:LambdaAction2} from the above eq.~\eqref{eq:actEmod} requires some care in picking both the generalised metric and parametrising the coset representative $m(x)$ in terms of the field $g(x)\in G$ of the $\lambda$-model.   One can choose $m(x) = \{ \bar{g} , \bar{g}^{-1} \}$ such that $\bar{g}^2 = g$ and the generalised metric is given in terms of the Killing metric $\kappa$ as
 \be
 {\cal H}_{AB} = \begin{pmatrix}  \epsilon^{-\frac{1}{2}}\kappa^{-1} & 0 \\ 0 &  \epsilon^{\frac{1}{2}}\kappa \end{pmatrix} \ , \quad   \epsilon^{\frac{1}{2}} = \frac{1- \lambda}{1+\lambda} \ . 
 \ee

\subsection{The $\eta$--$\lambda$ connection}
 
To complete the lectures let us return to the $\eta$-model on group manifolds and point out a feature that will close the circle of ideas presented. 
 Recall that the $\eta$-model takes exactly the form of a PL sigma model with    $ E_0^{-1} = \eta^{-1} \mathbbm{1}  - {\cal R} $ and hence can be T-dualised in this fashion.
 The relevant Drinfeld double was $G^\mathbb{C} = G \cdot AN $ and hence the PL T-dual to the  $\eta$-model takes the Borelian $AN$ group as its target space.  This theory, which we can call the $\lambda^\star$-model is in turn related to the $\lambda$-model by a subsequent analytic continuation of the geometry and the parameters that define the theory.   This connection, already somewhat anticipated in  \cite{Klimcik:2002zj,Vicedo:2015pna}, was shown explicitly for examples in \cite{Hoare:2015gda}  and then in \cite{Sfetsos:2015nya} before being confirmed in generality \cite{Klimcik:2015gba}.  Since this is a nice way to conclude the notes we give an explicit example based on the $\eta$ and $\lambda$ models for $SU(2)$ providing some detail. 

Let us begin with the $\eta$-model realised as PL sigma model for the double $SU(2)^\mathbb{C} = SU(2) \cdot E_3$ with the conventions detailed in appendix \ref{sec:AppDrinfeld}.
The $\eta$-model has the target space metric of a squashed $S^3$ given by,  in terms of right-invariant forms, 
\be
ds^2 = \frac{1}{t} \frac{1}{1+ \eta^2} \left( R_1^2 + R_2^2 + (1+\eta^2) R_3^2 \right) \, , 
\ee
together with a pure gauge two-form such that $H=d B=0$.    Cast in terms of the left invariant forms this geometry is equivalent to the PL sigma model with $E_0^{-1}= \eta^{-1} \mathbbm{1} + {\cal R}$.   The PL T-dual of this has $E_3$ as its target space and is described by the sigma model 
 \be
\tilde{S}_{\tilde{G}}[\tilde{g}] = \frac{1}{\eta t} \int \tilde{L}_{+ a}  \left[ \left(E_0 + \tilde\Pi[\tilde{g}] \right)^{-1} \right]^{ab}\tilde{L}_{- b}   \, .  
\ee
in which we restored the overall normalisation factor.  With the parameterisation of $E_3$ elements as 
\be
\tilde{g} =  \left(\begin{array}{cc} \mathrm{e}^{\chi/2}  & \mathrm{e}^{-\chi/2} r e^{- i \theta } \\ 0 &  \mathrm{e}^{-  \chi/2}    \end{array}\right) \ , 
\ee
and the results of appendix \ref{sec:AppDrinfeld} for $\tilde{\Pi}$ and the left-invariant forms we have the T-dual metric and B-field are given by 
\be
\begin{aligned}
 \widetilde{ds^2} =  &  \frac{1}{\Omega  \eta ^2 t}   \left[   \left(\Omega -4 r^2 e^{2 \chi }\right) d\chi^2  +4 \eta ^2 r^2 e^{2 \chi
   } d\theta^2+4  \left(\eta ^2 e^{2 \chi }+\left(\eta ^2+1\right)  r^2 
 \right)dr^2  \right. \\ 
   & \left.~~~~~~~~~~~~~~~~ -4 r \left(\left(\eta ^2-1\right) e^{2 \chi
   }+\left(\eta ^2+1\right) \left(r^2+1\right)\right)  dr d\chi    \right] \, , \\
   \widetilde{B} =  &  \frac{1}{\Omega  \eta  t} \left[2 r \left(\left(\left(\eta ^2-1\right) e^{2 \chi }+\left(\eta ^2+1\right)
   \left(r^2+1\right)\right) dr \wedge d \theta -2 r e^{2 \chi } d\theta
   \wedge d \chi  \right)     \right] \, , 
   \end{aligned}  \ee
   with 
   \be  
    \Omega = \left(\eta ^2+1\right) e^{4 \chi }+2 e^{2 \chi } \left(\eta ^2+\left(\eta
   ^2+1\right) r^2-1\right)+\left(\eta ^2+1\right) \left(r^2+1\right)^2  \, . 
  \ee
This is not quite the $\lambda$-theory; it needs to be supplemented by analytic continuation  \cite{Hoare:2015gda} of some of the angles defining the group element\footnote{In general this is the implementation  \cite{Klimcik:2015gba} that takes one from the Drinfeld double $\frak{g}^{\mathbb{C}} = \frak{g} + \frak{g}_{{\cal R}}$ of the $\eta$-model to the quasi Manin triple $\frak{d} = \frak{g} \oplus \frak{g}$ of the $\lambda$-model.  } 
  \be
  r \rightarrow i \sin \beta \sin \alpha \ , \quad  e^\chi \rightarrow \cos \alpha + i \sin \alpha \cos\beta \, ,
  \ee
   and also a relation between the deformation parameters and tensions of the models,
\begin{equation}
 \eta \rightarrow \frac{i (1- \lambda)}{(1+\lambda)}\ , \quad t \rightarrow \frac{\pi (1+\lambda)}{ k(1-\lambda)} \ . 
\end{equation}
Doing this one recovers exactly the metric and three form, $H$, of the $\lambda$-model of eq.~\eqref{su2def}.\footnote{With the $\tilde{B}$ given as above the Wick rotation results in a complex two-form whose imaginary part is pure gauge so that the field strength is real and matches that of the $\lambda$-model. }    Notice that acting on the parameter $q$ we have, 
\begin{equation}
q = e^{\eta t  } \leftrightarrow q = e^{\frac{i\pi}{k}} \, , 
\end{equation}
showing  that  $q$ real gets mapped to $q$ a root-of-unity. 

\section{Conclusion}

This then completes a voyage that has come full circle.   We saw how the common notion of  Abelian  T-duality has exotic variations of non-Abelian and Poisson-Lie T-duality.  Regardless of their uncertain nature as true quantum symmetries, both of these have great utility particularly within the context of holography and in connection to integrable models.  The $\eta$-deformation of the Principal Chiral Model  on a group manifold provides an integrable exemplar of Poisson-Lie dualisable theories, and after an analytic continuation its PL dual matches the $\lambda$-deformation.  This $\lambda$-deformation in turn interpolates between a CFT (WZW model) and the non-Abelian T-dual of the PCM.  

I hope that this introduction might motivate others to become involved in this nice confluence of ideas in integrablity, duality and holography.  There are many open directions and so I close with a few I find rather interesting:   
 \begin{itemize}
 \item Can we now, equipped with a large arsenal of tools, resolve the quantum nature of these dualities?  Can we understand the structure of $\alpha^\prime$ corrections?  Can we establish e.g. modular invariance of deformed theories? 
\item  Here we find an assortment of integrable $\sigma$-models with various embellishments.  In all cases   a serious visionary leap was required to come up with them, and it begs the question of what is the universal feature of integrable $\sigma$-models?  How can we understand and classify the full landscape of integrable two-dimensional theories? 
\item Recent work has explored $T \bar{T}$ \cite{Smirnov:2016lqw,Cavaglia:2016oda} and $J \bar{T}$ \cite{Guica:2017lia} deformations  which seem to share some features in common with the integrable deformations discussed.  Can a connection be made more precise (and see \cite{Araujo:2018rho} for recent work in this direction)?
\item What is the holographic understanding of the $\eta$- and $\lambda$-deformation at the level of the gauge theory? 
\item It is likely that these integrable deformations could show up in other contexts from Chern-Simons theories (see \cite{Schmidtt:2018hop}) to e.g. matrix model duals to blackholes.  How can we exploit these?   
\item The majority of this talk was in the context of type II string theory.  How then do these ideas fit within the M-theory umbrella?
\item In the motivation we began by speaking of the universal nature of duality across physics and not just in string theory.  One might then ask if there are exotic notions of duality such as PL that have applicability and utility on the lattice,  in spin chains or more generally in condensed matter systems?
\end{itemize}

\acknowledgments
  DCT is supported by a Royal Society University Research Fellowship {\em Generalised Dualities in String Theory and Holography} URF 150185 and in part by STFC grant ST/P00055X/1 and in part by the ``FWO-Vlaanderen'' through the project G006119N and by the Vrije Universiteit Brussel through the Strategic Research Program ``High-Energy Physics''.   DCT would like to thank the fellow organisers of the Corfu Institute 2018 -- with special thanks in particular to Athanasios Chatzistavrakidis --   and to G Zoupanos and the Cost Action COST Action MP1405 QSPACE for providing support to this meeting,  and to the participants for their enthusiasm,  excellent talks and questions.   Thanks to  the many colleagues and  collaborators whose contributions to this subject have been essential,   and to 
  S Demulder, S Driezen and G Piccinini for careful reading of this manuscript  and Y~ Lozano, E~\'{O}~Colg\'{a}in, C~N\'{u}\~{n}ez, K Sfetsos for comments on the draft. 
  
  \appendix

\section{Modified Supergravity}
\label{sec:mSUGRA}
Schematically the equations of modified supergravity \cite{Arutyunov:2015mqj,Wulff:2016tju} take the form
\begin{equation}
\begin{aligned}
 0=  &R_{mn}   -\frac{1}{4}H_{mpq}H_{n}{}^{pq} + \nabla_m X_n + \nabla_n X_m \\
&\qquad -   \Big( \frac{1}{2}({{\cal F}_1}^2)_{mn}+\frac{1}{4}({{\cal F}_3}^2)_{mn} +\frac{1}{96}({{\cal F}_5}^2)_{mn} - \frac{1}{4}g_{mn} \big( {\cal F}_1^2 +\frac{1}{6}{\cal F}_{3}^2 \big)\Big)  \ , \\
0  =  &d   \star H  + {\cal F}_1\wedge \star {\cal F}_3 + {\cal F}_3 \wedge {\cal F}_5  -2 \star d X -2 X \wedge \star H \ , \\
0  =   &R +4 \nabla_n X^n -  4 X_n X^n - \frac{1}{12} H^2 \, ,   \\
0  = &  {\bf d} {\cal F} \equiv (d+ H\wedge - Z\wedge - \iota_I ) {\cal  F}      \ . 
\end{aligned}
\end{equation}
 Here the vector $X$ is given by 
\begin{equation}
X= Z+I \ , 
\end{equation}
with the constraints
\begin{equation}
dZ + \iota_I H = 0  \ , \quad \iota_I Z = 0  \ , \quad L_I g = L_I H = 0 \ . 
\end{equation}
For the case of $I=0$ we have that $X= d\phi$ and the conventional supergravity is recovered with  ${\cal F}  = e^\phi  \mathbb{F}$ the RR poly form.    In general we identify the ``dilaton'' as the exact piece of $Z$;
\begin{equation}
Z= d\phi + \iota_I B - V \ , \quad L_I B = d V \ . 
\end{equation}

\section{Gaiotto-Maldacena Geometries} 
The  backgrounds dual to the SCFTs   of  \cite{Gaiotto:2009we}  were given by  Gaiotto and Maldacena in
\cite{Gaiotto:2009gz}. The solutions we are interested in are found in Type IIA supergravity and can be   completely specified by a potential function $V(\sigma,\eta)$. 
Denoting
\begin{equation}
\dot{V}=\sigma \partial_\sigma V, \;\;\;\ddot{V}= \sigma^2\partial^2_\sigma V+\sigma \partial_\sigma V;\;\;\; V'=\partial_\eta V, \;\;\; V''=\partial^2_\eta V,\nonumber
\end{equation}
one can write the Type IIA 
background  (in string frame) as
\begin{eqnarray}
& & ds^2=\alpha'(\frac{2\dot{V} -\ddot {V}}{V''})^{1/2}
\Big[  4 AdS_5 +\mu^2\frac{2V'' \dot{V}}{\Delta} 
{d \Omega^{2}_2(\chi,\xi)}+\mu^2\frac{2V''}{\dot{V}}  
(d\sigma^2+d\eta^2)+ \mu^2\frac{4V'' \sigma^2}{2\dot{V}-\ddot{V}} 
d{\beta}^2 \Big], \nonumber\\
& & A_1=2\mu^4\sqrt{\alpha'}
\frac{2 \dot{V} \dot{V'}}{2\dot{V}-\ddot{V}}d{\beta},\;\;\;\; 
e^{4\phi}= 4\frac{(2\dot{V}-\ddot{V})^3}{\mu^{4}V'' \dot{V}^2 \Delta^2}, 
\quad {\Delta = (2 \dot{V} - \ddot{V}) V'' + (\dot{V}')^2} \ ,  \nonumber \\
& & B_2=2\mu^2\alpha' (\frac{\dot{V} \dot{V'}}{\Delta} -\eta) 
d\Omega_2,\;\;\; {C}_3={-} 4\mu^4 \alpha'^{3/2}
\frac{\dot{V}^2 V''}{\Delta}d{\beta} \wedge d\Omega_2.
\label{metrica}
\end{eqnarray}
The radius of the space is $\mu^2\alpha'=L^2$. We use 
the  two-sphere metric $d \Omega^{2}_2(\chi,\xi)=d\chi^2+\sin^2\chi d\xi^2$, with corresponding volume 
form $d\Omega_{2}= \sin\chi d\chi \wedge d\xi$. The usual definition
$F_4= dC_3 + A_1\wedge H_3$ is also used.

To write backgrounds in this family, one should find the function $V(\sigma,\eta)$ that solves
a Laplace problem with a given charge density $ \lambda (\eta)$ and boundary conditions,
\begin{eqnarray}
& & \partial_\sigma[\sigma \partial_\sigma V]+\sigma \partial^2_\eta V=0,\nonumber\\
& &  \lambda (\eta)= \sigma\partial_\sigma V(\sigma,\eta)|_{\sigma=0}, \;\;\;\;  \lambda (\eta=0)=0,\;\;  \lambda (\eta=N_c)=0.
\label{ecuagm1}
\end{eqnarray}
The boundary condition at $\eta = N_c$ ensures that the corresponding SCFT quiver will have finite length though in both the Sfetsos-Thompson and  Maldacena-Nunez solutions this requirement is not satisfied.

   \section{Maillet $r-s$ form and conserved charges}
   \label{app:Maillet} 

A useful treatment of the conserved charges in integrable systems comes from the Maillet algebra \cite{Maillet:1985ek}, that governs the Poisson bracket of the spatial component of the $\frak{g}^{\mathbb{C}}$ valued Lax ${\cal L}(z) \equiv {\cal L}_+(z) - {\cal L}_-(z)$.  We allow this Lax to act in a tensor product with notation ${\cal L}_{1} (z) = {\cal L}(z) \otimes \mathbbm{1} $ and ${\cal L}_{2} (z) = \mathbbm{1}  \otimes {\cal L}(z)  $.  The general form of the algebra reads 
\be
\begin{aligned}
\{ {\cal L}_1(\sigma_1,z) , {\cal L}_2(\sigma_2, w)   \}_{P.B.} =& [ r_{12}(z,w) , {\cal L}_1(\sigma_1,z)  + {\cal L}_2(\sigma_1, w)  ] \delta(\sigma_1-\sigma_2) \\
& - [ s_{12}(z,w) , {\cal L}_1(\sigma_1,z)  - {\cal L}_2(\sigma_1, w)  ] \delta(\sigma_1-\sigma_2) - 2 s_{12}(z,w) \delta'(\sigma_1-\sigma_2) \, .
\end{aligned}
\ee
The derivative of the delta function in this algebra signifies this theory is non-ultra-local.   Often we can express the operators $r_{12}$ and $s_{12}$ in terms of the Casimir $t_{12} = \sum_a T_a \otimes T_a$ with the spectral parameter dependence captured by the so-called twist function $\phi(z)$:
\be
r_{12}(z,w) = \frac{1}{z- w}\left(\phi^{-1}(z)+  \phi^{-1}(w)\right)t_{12} \ , \quad s_{12}(z,w) = \frac{1}{z- w}\left(\phi^{-1}(z)-  \phi^{-1}(w)\right)t_{12} \, . 
\ee 
We wish to expand the monodromy matrix $T(z) = P\exp \int  {\cal L}(z)$ around special points $z_\star$   which exhibit the algebra of charges (e.g. Yangian or quantum group).  These special points are ones for which $r(z,w)$ has a finite limit as $z \to z_\star$ and $w \to z_\star$.  When expressible in terms of the twist function as above, the special points occur at the poles of the twist function (see \cite{Vicedo:2015pna} for a detailed treatment).

As an example, for the $\eta$-deformation on a group we have that 
\be
\phi(z) = \frac{1+\eta^2}{2  t} \frac{1-z^2}{\eta^2 + z^2} \Rightarrow z_\star = \pm i \eta \ .
\ee
We can now review the construction of conserved charges \cite{Kawaguchi:2011pf,Delduc:2013fga} realising the $G_L$ quantum symmetry for the $SU(2)$ $\eta$-deformation.  Here we use $\frak{su}(2)$ generators   $[ T_\pm ,T_3 ] =\pm  i T_\pm$, $[T_+ , T_- ]=   i T_3$ and  define  $ g^{-1}dg  \equiv u^+ T_+ + u^- T_- + u^3 T_3$.  We introduce  some  currents,
\be
j_\pm =  \frac{i}{t (1+\eta^2) }  \left(\eta\, u_{\sigma}^\pm  \mp  i u_{\tau}^{\pm }  \right) \,, \quad j_3= - \frac{1}{  \eta t  } u_{\tau}^3   \, , 
\ee
 which obey  simple Poisson brackets relations:
\be
\begin{aligned}
\{ j_3 (\sigma_1) , j_3(\sigma_2) \} &=  0  \ , \\
\{ j_\pm (\sigma_1) , j_3(\sigma_2) \} &= \pm i j_\pm (\sigma_2) \delta(\sigma_1- \sigma_2) \ ,  \\ 
\{ j_+ (\sigma_1) , j_- (\sigma_2) \} &=   i  j_3(\sigma_2) \delta(\sigma_1- \sigma_2)  \ . 
 \end{aligned}
\ee
The Lax of this theory was naturally formulated in terms of right invariant one-forms and in order to see the $G_L$ symmetry we should consider the monodromy built from the gauge transformed Lax ${\cal L}^g(z) = g^{-1}  {\cal L}(z) g - g^{-1} \partial_\sigma g$.  If we expand this around the poles in  the twist function we find     
\be
{\cal L}^g(z= \pm  i \eta)   =    i  t  \eta \left( -2 j_\pm T_\pm   \pm j_3 T_3  \right) \ .
\ee
Using the fact that the Cartan element can be factored in the path ordered exponential occurring in the monodromy matrix \cite{Kawaguchi:2012gp} one is led to construct  (non-local) currents 
\be
\begin{aligned}
\frak{J}_+(\sigma ,\tau) &= j_+(\sigma ,\tau) \exp\left[t \eta  \int_{\sigma}^\infty  j_3 (\hat{\sigma}  ,\tau)  d\hat{\sigma}  \right] \ , \\  
\frak{J}_-(\sigma ,\tau) &= j_-(\sigma ,\tau) \exp\left[ - t \eta   \Sigma \int_{-\infty}^\sigma  j_3 (\hat{\sigma}  ,\tau)  d\hat{\sigma}  \right] \ , \\  
\frak{J}_3(\sigma ,\tau) &= j_3(\sigma ,\tau) \ . 
 \end{aligned}
\ee 
The equations of motion imply $\partial_\tau \frak{J} = \partial_\sigma \tilde{ \frak{J}}$ for some $\tilde{ \frak{J}}$ whose explicit form is not important to us and thus that the charges $\frak{Q} = \int_{-\infty}^\infty \frak{J} d\sigma$   are conserved subject to standard boundary fall off.  The Poisson brackets give  
\be\begin{aligned}
\{ \frak{J}_+(\sigma_1), \frak{J}_-(\sigma_2) \} 
&=&   \frac{i}{ 2 t \eta } \, \delta(\sigma_1- \sigma_2) \,    \partial_{\sigma_2}    \exp\left[ t\eta  \left(\int_{\sigma_2}^\infty  -  \int_{-\infty}^{\sigma_2}\right)  j_3 (\hat{\sigma}) d\hat{\sigma}  \right]  \ ,
 \end{aligned} 
\ee
 where we note that ``cross terms'' involving the non-local exponentials cancel.   Thus one finds that, with suitable normalisation, 
 \be\label{eq:quantumgroup}
 \{ \frak{Q}_+, \frak{Q}_- \}  =   i\, \frac{q^{\frak{Q}_3} - q^{- \frak{Q}_3}  }{q -q^{-1} } \ , \quad  \{ \frak{Q}_\pm, \frak{Q}_3 \}  =  \pm i \,\frak{Q}_\pm \ , \quad  
 q =  e^{ t \eta } \ .
 \ee

 \section{The Drinfeld Double $\frak{su}(2) \oplus \frak{e}_3$ }
\label{sec:AppDrinfeld}

We follow the parametrisation of \cite{Sfetsos:1999zm}. We choose  a   representation of  $\frak{su}(2)$ and  $\frak{e}_3$  generators given in terms of Pauli matrices by block diagonal matrices 
\def\s{\sigma}
\def\diag{{\textrm{diag}}}
 \begin{equation*}
 \begin{split}
&T_1= \frac{1}{2}\diag(\s^1 , \s^1) \ , \quad T_2 = \frac{1}{2}\diag(\s^2 , \s^2) \ , \quad T_3=  \frac{1}{2}\diag(\s^3 , \s^3) \ ,  \\
&\tilde{T}^1 = i \diag( \s^+ , -\s^-) \ , \quad \tilde T^2 = \diag(\sigma^+  , \sigma^-)  \ , \quad \tilde{T}^3 = \frac{i}{2} \diag( \s_3 , - \s_3) \ ,
\end{split}
 \end{equation*}
 where $\s^\pm = \frac{1}{2} (\s^1 \pm i \s^2)$.   We define an inner product on $\frak{su}(2) \oplus \frak{e}_3$ by
  \begin{equation*}
 \langle\langle X , Y \rangle\rangle  = -i\,  \mathrm{tr} \left(P_u X P_u Y - P_d X P_d Y \right) \ ,
  \end{equation*}
 where $P_u$ projects onto the top-left two-by-two block and $P_d$ onto the bottom-right.  If we let $T^{A} = \{ T_{a} , \tilde{T}^{b} \}$ with $A = 1\dots 6$, $a,b= 1,2,3$, be a basis for the generators of the double then
 \begin{equation*}
 \langle\langle  T_{A}, T_{B}  \rangle\rangle = \eta_{AB} =   \left( \begin{array}{cc} 0 & \mathbbm{1}  \\ \mathbbm{1}  & 0 \end{array} \right) \ ,
 \end{equation*}
  indicating that $\frak{su}(2)$ and $\frak{e}_{3}$ span mutually orthogonal maximal isotropic subspaces with respect to this inner product.      We   parametrise a group element as
 \begin{equation*}
 \begin{aligned}
 & g_0 = \exp(i/2 \varphi \s^3) \exp(i/2 \theta \s^2) \exp(i/2 \psi \s^3) \ , \quad  g_{SU(2)} =  \diag(g_0  ,g_0 ) \ ,  \\
  & g_{+} = \left(\begin{array}{cc} \mathrm{e}^{\chi/2}  & \mathrm{e}^{-\chi/2}(y_1 - i y_2) \\ 0 &  \mathrm{e}^{-  \chi/2}    \end{array}\right) \ , \quad
    g_{-} = \left(\begin{array}{cc} \mathrm{e}^{-\chi/2}  & 0  \\  - \mathrm{e}^{-\chi/2}(y_1 +  i y_2)  &   \mathrm{e}^{   \chi/2}    \end{array}\right)
     \ , \quad g_{E_3} = \diag( g_+, g_-)\,.
    \end{aligned}
 \end{equation*}
and further define $y_1 + i y_2  = r e^{i \vartheta}$.

Using this parameterisation one finds that the left-invariant and right-invariant one-forms for $\frak{su}(2)$  are given by
 \begin{equation*}
 \begin{aligned}
& L^1 =  \sin \theta  \cos \psi   \mathrm{d} \varphi  -  \sin \psi  \mathrm{d} \theta   \ , \quad
 L^2 =\sin \theta  \sin \psi
\mathrm{d} \varphi  + \cos \psi    \mathrm{d} \theta\ ,  \quad
  L^3 =  \cos \theta  \mathrm{d} \varphi  + \mathrm{d} \psi \ , \\ 
  &R^1 = - \sin \theta  \cos \phi   \mathrm{d} \psi  +  \sin \phi  \mathrm{d} \theta   \ , \quad
 R^2 =\sin \theta  \sin \phi
\mathrm{d} \psi+ \cos \phi    \mathrm{d} \theta\ ,  \quad
  R^3 =  \cos \theta  \mathrm{d} \psi  + \mathrm{d} \phi \ , \\ 
    \end{aligned}
 \end{equation*}
whilst those of  $\frak{e}_{3}$   by  
 \begin{equation*}
 \begin{aligned}
  & \tilde{L}_1 = -e^{-\chi }  \mathrm{d} y_1\ , \quad   \tilde{L}_2 = -e^{-\chi }  \mathrm{d} y_2 \ ,
\quad     \tilde{L}_3 =  - \mathrm{d} \chi\, , \\
 & \tilde{R}_1 =  - \mathrm{d} y_1 + y_1  \mathrm{d} \chi  \ , \quad   \tilde{R}_2 =   - \mathrm{d} y_2 + y_2  \mathrm{d} \chi   \ ,
\quad     \tilde{R}_3 =  - \mathrm{d} \chi\, . 
    \end{aligned}
 \end{equation*}
The combination of adjoint actions defined in \eqref{eq:pimat} are given by 
\begin{equation*}
\begin{split}
&\Pi^{ab} = \left(
\begin{array}{ccc}
 0 & -2 \sin ^2\frac{\theta }{2} & -\sin \theta  \sin \psi  \\
 2 \sin ^2\frac{\theta }{2} & 0 & \cos \psi  \sin \theta  \\
 \sin \theta  \sin \psi  & -\cos \psi  \sin \theta  & 0 \\
\end{array}
\right) \ , \\
\tilde{\Pi}_{ab} &=
\left(
\begin{array}{ccc}
 0 & \frac{1}{2} e^{-2 \chi } \left(-r^2+e^{2 \chi }-1\right) &
   -e^{-\chi } y_2 \\
 \frac{1}{2} e^{-2 \chi } \left(r^2-e^{2 \chi } +1\right) & 0 &
   e^{-\chi } y_1 \\
 e^{-\chi } y_2 & -e^{-\chi }y_1 & 0 \\
\end{array}
\right) \ .
\end{split}
\end{equation*}

\bibliography{literatur}  
\bibliographystyle{JHEP}
%
%
\end{document}